\newcommand{\eps}{\epsilon}

\newcommand{\dsv}{\Delta^{sv}}
\newcommand{\stm}{$S$-$T_-$}
\newcommand{\stz}{$S$-$T_{0}$}
\newcommand{\dez}{\Delta E_Z}
\newcommand{\ez}{E_Z}
\newcommand{\tm}{T_-}

\newcommand{\UD}{\ket{\uparrow^{\plus}\downarrow^{\minus}}} 
\newcommand{\DU}{\ket{\downarrow^{\plus}\uparrow^{\minus}}}
\newcommand{\minus}{\scalebox{0.6}{$-$}}
\newcommand{\plus}{\scalebox{0.6}{$+$}}

\newcommand{\U}{\uparrow}
\newcommand{\D}{\downarrow}

\documentclass[%
reprint,amsmath,amssymb,aps,prl,superscriptaddress,
]{revtex4-1}

\usepackage{graphicx}
\usepackage{dcolumn}
\usepackage{bm}
\usepackage{epstopdf}
\usepackage{float}
\usepackage{braket}
\usepackage{dsfont}
\usepackage{amsmath}
\usepackage{xcolor}
\usepackage{lipsum}
\usepackage{color,soul}

\graphicspath{{figures/}}

\begin{document}

\title{Coherent spin-valley oscillations in silicon}

\author{Xinxin Cai}
\thanks{These authors contributed equally.}

\author{Elliot J. Connors}
\thanks{These authors contributed equally.}

\affiliation{Department of Physics and Astronomy, University of Rochester, Rochester, NY, 14627 USA}


\author{John M. Nichol}
\email{john.nichol@rochester.edu}

\affiliation{Department of Physics and Astronomy, University of Rochester, Rochester, NY, 14627 USA}

\begin{abstract}
Electron spins in silicon quantum dots are excellent qubits because they have long coherence times, high gate fidelities, and are compatible with advanced semiconductor manufacturing techniques~\cite{Watson2018,Mi2018,xue_benchmarking_2019,Huang2019,Andrews2019Triple,zwerver_qubits_2021,noiri2021fast,xue2021computing}. The valley degree of freedom, which results from the specific character of the Si band structure, is a unique feature of electrons in Si spin qubits~\cite{ando_electronic_1982,Zwanenburg2013}. However, the small difference in energy between different valley levels often poses a challenge for quantum computing in Si~\cite{Friesen2007valley,culcer2010quantum}. Here, we show that the spin-valley coupling in Si, which enables transitions between states with different spin and valley quantum numbers~\cite{Yang2013,Hao2014,huang_spin_2014,zhang2020anisotropy}, enables coherent control of electron spins in Si. We demonstrate coherent manipulation of effective single- and two-electron spin states in a Si/SiGe double quantum dot without ac magnetic or electric fields. Our results illustrate that the valley degree of freedom, which is often regarded as an inconvenience, can itself enable quantum manipulation of electron spin states.

\end{abstract}


\maketitle

\section{Introduction}


Recent demonstrations of gate fidelities approaching the regime of fault-tolerance~\cite{noiri2021fast,xue2021computing,madzik2021precision,mills2021two} confirm the potential of quantum computing in Si. Compared with other solid-state platforms, Si qubits offer the promise of extremely long coherence times and compatibility with large-scale semiconductor fabrication. Despite this promise, quantum computing with electron spins in Si must contend with a major complication:
energy levels in Si quantum wells or interfaces are nominally two-fold degenerate. The conduction band minima in Si occur near six equivalent faces of the first Brillouin zone~\cite{schaffler1997}, and in quantum wells, tensile strain associated with the semiconductor growth raises the energies of four of the six levels.
The remaining two low-lying ``valleys" are approximately degenerate, and the splitting between these levels depends on the microscopic details of the semiconductor~\cite{ando_electronic_1982,Zwanenburg2013}. 

Because of this fact, electron spins in Si quantum dots possess not only spin and orbital quantum numbers, but also valley quantum numbers. 
Unfortunately, the energy difference~\cite{Friesen2007valley} between otherwise identical quantum states with different valley quantum numbers, which is often in the range of $10^{-5}$~eV, presents various obstacles for quantum information processing with electrons in Si.
Small valley splittings can, for example, reduce control fidelities~\cite{Kawakami2014}, limit initialization and readout of spin qubits~\cite{culcer2010quantum,jones_spin-blockade_2019}, and hamper the operation of qubits whose Hamiltonians depend on the valley splitting~\cite{penthorn2019two,mi_landau-zener_2018}.
Measuring valley splittings~\cite{Borselli2011,mi_high-resolution_2017,chen_detuning_2021} and increasing them through fabrication~\cite{sasaki2009well,Chen2021,mcjunkin2021valley,neyens2018critical} or device-operation approaches~\cite{goswami2007controllable,Yang2013} are major areas of research.
Furthermore, the valley and spin degrees of freedom can also interact through spin-orbit coupling.
Such spin-valley coupling can introduce a ``hot spot" with enhanced relaxation between single-spin states when the Zeeman energy equals the valley splitting~\cite{Yang2013,Hao2014,huang_spin_2014,zhang2020anisotropy,hollmann_large_2020}.

Because of these challenges, the valley quantum number in Si is often not used and frequently viewed as a defect or nuisance rather than a feature, though some types of qubits take advantage of the valley splitting in Si~\cite{schoenfield2017coherent,penthorn2019two,Koh2012}. In this work, we show that the valley degree of freedom and the associated spin-orbit-driven coupling between states with different spin and valley quantum numbers~\cite{Yang2013,Hao2014,huang_spin_2014,zhang2020anisotropy} can enable coherent electron-spin rotations between different effective single- and two-electron spin states. This coupling between spin states is simple to use because it does not require ac magnetic or electric fields. We also show that spin-valley coupling enables exploring previously-inaccessible spin-spin couplings, including singlet-triplet and triplet-triplet transitions, revealing a family of new spin qubits. This method of controlling spin qubits takes advantage of the unique nature of the Si band structure and offers an especially simple method for coherent manipulation of single- and multi-spin states in Si. 

\begin{figure}[t]
	\includegraphics{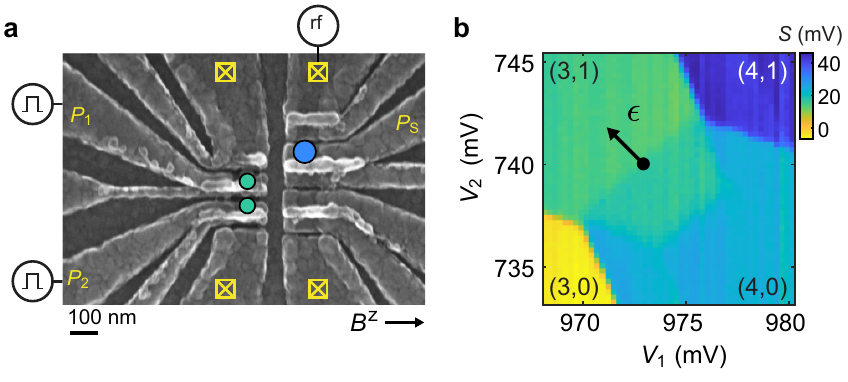}
	\caption{Experimental setup.
		\textbf{a}, Scanning electron micrograph of a device that is nominally identical to the one used. A double dot is formed under plunger gates $P_1$ and $P_2$. A sensor dot is formed beneath plunger gate $P_S$. Yellow boxes with an ``x" in them represent Ohmic contacts to the quantum well. The direction of the external magnetic field $B^z$ is indicated.
		\textbf{b}, Pauli spin blockade at the (3,1)-(4,0) charge transition. The trapezoid in the (4,0) charge region indicates the Pauli spin blockade region.  $S$ is the measured charge-sensor signal. 
		The detuning $\epsilon$ is defined as a change in the plunger gate voltages along the diagonal of the charge stability diagram such that $(\Delta V_{1},\Delta V_{2}) = (-\epsilon,\epsilon)$, with $\epsilon=0$ at the (4,0)-(3,1) charge transition.}\label{fig:device}
\end{figure}

\section{Device and Hamiltonian}
We employ a double quantum dot fabricated in an undoped Si/SiGe heterostructure with an overlapping-gate architecture \cite{Zajac2015,Angus2007,Zajac2015,Borselli2015,Zajac2016,Connors2021charge} (Fig.~\ref{fig:device}a). Quantum dots 1 and 2, which together form the double dot, are formed beneath plunger gates $P_1$ and $P_2$, respectively, with their electron occupancies controlled by the positive voltages applied to the gates, $V_1$ and $V_2$. 
We use the quantum dot beneath gate $P_S$ as a charge sensor and measure its conductance via rf reflectometry for fast spin-state readout \cite{Reilly2007FastRF,Barthel2009,Connors2020}. The device is cooled in a dilution refrigerator to a base temperature of about 10~mK.

\begin{figure}[t]
	\includegraphics{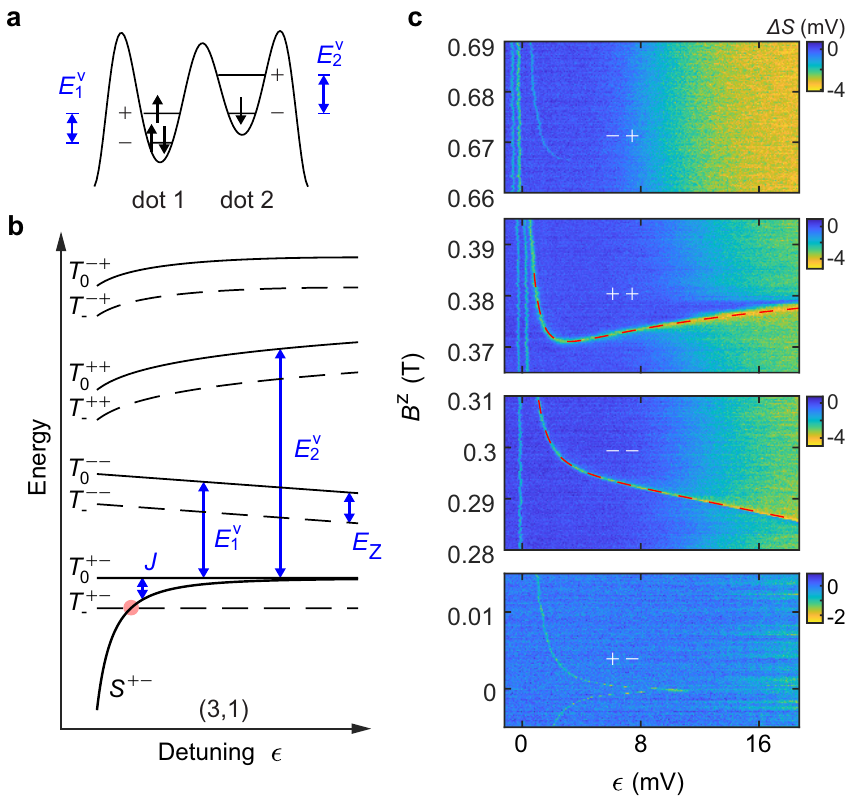}
	\caption{Spin funnels.
		\textbf{a}, Schematic of occupied levels associated with the $\ket{\U^{\plus} \D^{\minus}}$ state.
		\textbf{b}, Energy level diagram showing the relevant energy levels for the spin funel experiments. $\ez=\bar{g} \mu_B B^z$ is the average Zeeman energy of the two dots. The red circle denotes the mixing point between $\ket{S^{\plus\minus}}$ and $\ket{T_-^{\plus\minus}}$.
		\textbf{c}, The four spin funnels observed in this experiment.
		In each graph the valley configuration of the relevant $T_-$ state is indicated. 
		The dashed red lines for the second and third spin funnels show the extracted values used to calculate $E^v_{1(2)}$. 
		$\Delta S$ is the measured charge-sensor signal minus the most likely signal value occurring at $\epsilon<5$~mV (corresponding to the singlet signal value) in each plot.
	}\label{fig:funnel}
\end{figure}

We tune the device to the (4,0)-(3,1) charge transition, where the (i,j) notation indicates that dot 1(2) is occupied with i(j) electrons (Fig.~\ref{fig:device}b). We describe this system of four electrons using a phenomenological Hubbard model, in addition to valley, spin, and spin-valley terms in the total Hamiltonian: $H=H^{Hub}+H^v+H^s+H^{sv}$ (see Methods). Within this framework, the relevant four-electron quantum states consist of Slater determinants of the lowest four single-particle energy levels of the two quantum dots: two valleys ($-$ and $+$) in dot 1 and two valleys ($-$ and $+$) in dot 2. In the (4,0) charge configuration, both valleys in dot 1 are full. In the (3,1) charge configuration, two electrons in dot 1 fill a single valley, either $+$ or $-$. The remaining electron occupies the unfilled valley in dot 1, and the electron in dot 2 can occupy either of its valleys. Below, we refer to the unpaired electrons (those without another electron in the same dot and valley) as ``valence" electrons, and we label the four-spin states by the configuration of the valence electrons. Figure~\ref{fig:funnel}a, for example, depicts the state $\ket{\U^{\plus} \D^{\minus}}$, where the left arrow indicates the spin of the valence electron in dot 1, the right arrow indicates the spin of the valence electron in dot 2, and the superscripts indicate the valley of each spin.
As we show in the following, while, in general, the electron dynamics in our device involve all four electrons, viewing these four-electron spin states as effective two-electron states results in a surprisingly intuitive and useful picture of the system.

In the absence of spin-valley coupling, the low-energy states of this Hamiltonian yield an effective singlet-triplet qubit~\cite{Hu2001,Barnes2011Screening,vorojtsov2004spin,Nielsen2013,Harvey-Collard2017,Harvey-Collard2018,West2019,Connors2021charge,Jock2021Valley,seedhouse_pauli_2021,Liu2021Gradient}, whose eigenstates are $\ket{S^{\plus\minus}}= (\ket{\U^{\plus} \D^{\minus}}-\ket{\D^{\plus} \U^{\minus}})/\sqrt{2}$ and $\ket{T_0^{\plus\minus}}= (\ket{\U^{\plus} \D^{\minus}}+\ket{\D^{\plus} \U^{\minus}})/\sqrt{2}$. These states are separated in energy by the exchange coupling $J(\epsilon)$. In addition, if $J(\epsilon)< \dez=g_1 \mu_B B^z-g_2 \mu_B B^z$, where $g_{1(2)}$ is the $g$ factor of dot 1(2)~\cite{Liu2021Gradient,Connors2021charge} and $B^z$ is the externally applied in-plane magnetic field, the eigenstates of the singlet-triplet qubit are $(\ket{T_0^{\plus\minus}}+\ket{S^{\plus\minus}})/\sqrt{2}=\ket{\U^{\plus} \D^{\minus}}$ and $(\ket{T_0^{\plus\minus}}-\ket{S^{\plus\minus}})/\sqrt{2}=\ket{\D^{\plus} \U^{\minus}}$~\cite{Petta2005,Foletti2009}. 

All measurements presented in this paper begin with the state initialized as a singlet in the (4,0) charge configuration. We separate the valence electrons into the $\ket{S^{\plus\minus}}$ state by tunneling into the (3,1) configuration. Then, we manipulate the joint spin state of the valence electrons by pulsing $\epsilon$ (Fig.~\ref{fig:device}b). We define $\epsilon$ such that positive values of $\epsilon$ feature positive voltage pulses away from the idling tuning on $P_2$ and negative voltage pulses on $P_1$. To measure the device, we perform a Pauli spin blockade (PSB) readout~\cite{Petta2005,Foletti2009}, which allows us to distinguish between the singlet and triplet configurations of the two valence electrons (Fig.~\ref{fig:device}b). In the absence of interactions with other spin states, such as any of the polarized triplets, the two electrons remain with high probability in the subspace spanned by $\ket{S^{\plus\minus}}$ and $\ket{T_0^{\plus\minus}}$.

\section{Valley splittings}
We begin by extracting the valley splitting in each dot, $E^v_{1(2)}$, from a ``spin funnel" measurement~\cite{Petta2005,Liu2021Gradient}. Near zero magnetic field, a spin funnel reveals the field $B^z$ where the lowest-energy singlet, $\ket{S^{\plus\minus}}$, and the lowest-energy polarized triplet, $\ket{T_-^{\plus\minus}}= \ket{\D^{\plus} \D^{\minus}}$, come into resonance, which occurs when $\bar{g} \mu_B B^z = J(\epsilon)$, where $\bar{g}=(g_1+g_2)/2$ is the average $g$-factor of dots 1 and 2 (Fig.~\ref{fig:funnel}b). However, at higher magnetic fields, the excited $T_-$ states ($\ket{T_-^{\minus\minus}}=\ket{\D^{\minus} \D^{\minus}}$, $\ket{T_-^{\plus\plus}}=\ket{\D^{\plus} \D^{\plus}}$, or $\ket{T_-^{\minus\plus}}=\ket{\D^{\minus} \D^{\plus}}$) come into resonance with $\ket{S^{\plus\minus}}$ when  $\bar{g} \mu_B B^z = \Delta + J(\epsilon)$, where $\Delta$ can be any of $\{E^v_1, E^v_2, E^v_1+E^v_2\}$ (Fig.~\ref{fig:funnel}b).   


Figure~\ref{fig:funnel}c shows the four spin funnels we observe, which in the following we refer to as the ``first," ``second," ``third," and ``fourth" spin funnels according to the order in which they occur when sweeping the magnetic field from zero.
In each panel, the thin bright lines indicate the magnetic-field and detuning values with relatively high triplet-return probabilities, which occur when $\ket{S^{\plus\minus}}$ comes into resonance with one of the above mentioned triplet states.
We determine $J(\epsilon)$ from the first spin funnel, together with measurements of coherent exchange oscillations (see Methods).
We use the magnetic-field and detuning values of the second and third spin funnels to extract the valley splitting of dot 1 and dot 2, respectively. 

The data of Fig.~\ref{fig:funnel}c reveal two notable features. First, $E^v_1$ and $E^v_2$ clearly depend on the gate voltages, as evidenced by the downward and upward sloping lines of the second and third spin funnels at large values of $\epsilon$. 
We extract the voltage dependencies of the valley splitting of each dot by assuming that $E^v_1 = E^v_1(\Delta V_1)$ and $E^v_2 = E^v_2(\Delta V_2)$.
We find $dE^v_1/d\Delta V_{1}=0.066$~meV/V and $dE^v_2/d\Delta V_{2}$ varies from $0.033$ to $0.54$~meV/V across the voltage range that we measure.
The negative slope of the second spin funnel occurs because positive changes to $\eps$ involve a negative change to $\Delta V_1$. 
These values are in agreement with recent reports in Si quantum dots~\cite{Liu2021Gradient,Jock2021Valley}.

The second notable feature of the data in Fig.~\ref{fig:funnel}c is the second and third spin funnels are substantially brighter than the first, especially deep in (3,1). Usually, the dominant \stm{} mixing mechanism is a transverse hyperfine gradient~\cite{Stepanenko2012}, which is expected to be small in Si. The presence of a small transverse gradient, for example, explains why the first spin funnel is relatively dim. To understand the prominence of the second and third spin funnels, recall that deep in the (3,1), where $J(\epsilon) \approx 0$, the singlet-triplet-qubit eigenstates are $\ket{\U^{\plus} \D^{\minus}}$ and $\ket{\D^{\plus} \U^{\minus}}$. On the second spin funnel, the state $\ket{\D^{\minus} \D^{\minus}}$ comes into resonance with these two eigenstates. A mechanism that simultaneously flips the spin in dot 1 and changes its valley could induce a rapid transition between the triplet and one of the eigenstates.  Likewise, a mechanism that can simultaneously flip the spin and change the valley of the electron in dot 2 could generate a similar strong transition between $\ket{\D^{\plus}\D^{\plus}}$ and the opposite eigenstate.

\section{Single-Spin-valley coupling}
In Si, spin-orbit coupling can generate such a ``spin-valley" coupling between single-spin states with different spin and valley quantum numbers: $\braket{\U^{\minus}|H^{sv}|\D^{\plus}}=\dsv$~\cite{Yang2013,Hao2014,huang_spin_2014,zhang2020anisotropy}. Spin-valley couplings have been explored for their effects on spin-state energies~\cite{Hao2014,scarlino2017dressed,Veldhorst2015b}, spin relaxation times~\cite{Yang2013,Tahan2014,zhang2020anisotropy,hollmann_large_2020}, and as ways to assist with qubit manipulation~\cite{corna_electrically_2018,bourdet2018all,Jock2021Valley}. However, despite the prominent $S$-$T_-$ mixing observed here and previously~\cite{Liu2021Gradient}, which suggests substantial spin-valley coupling rates, this mechanism has not been explored as a method itself to coherently manipulate spin states.

\begin{figure}[t!]
	\includegraphics{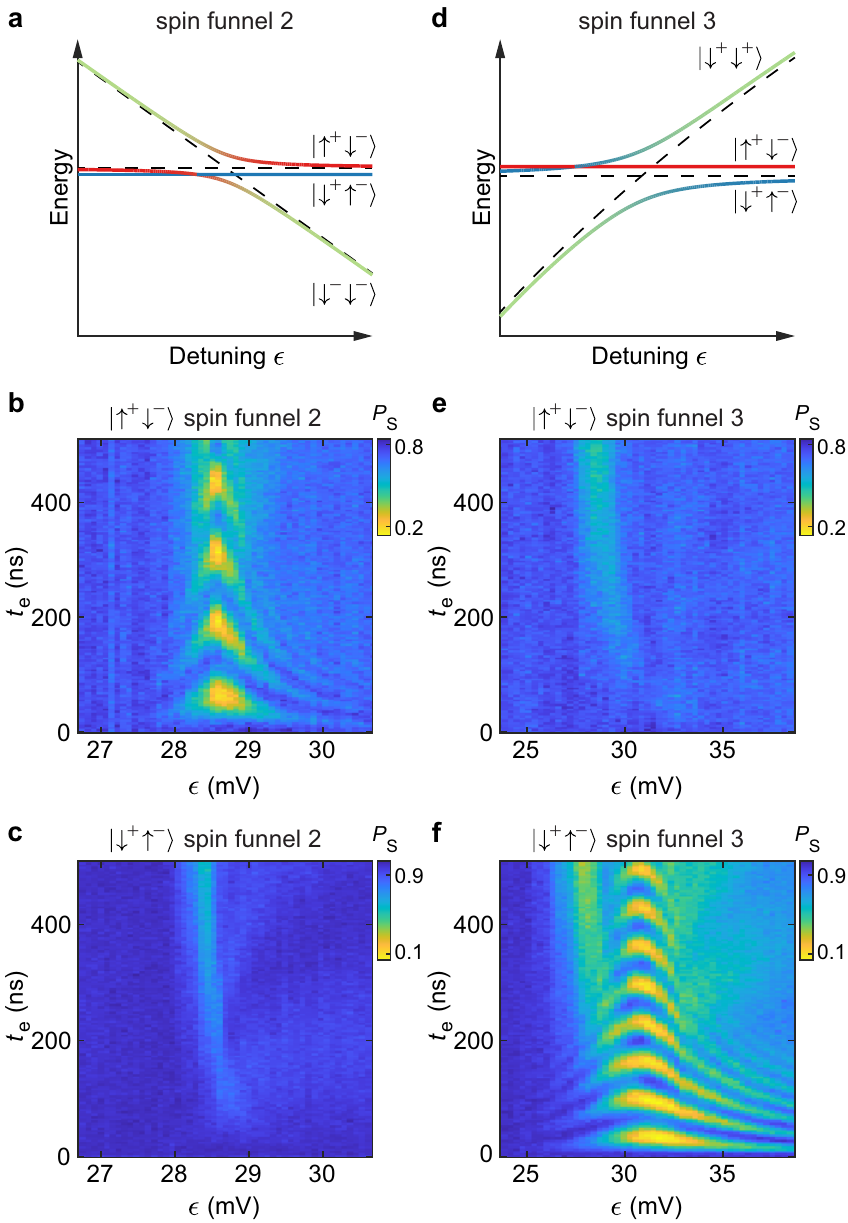}
	\caption{Single-spin Rabi oscillations driven by spin-valley coupling. 
	\textbf{a}, Energy level diagram at large $\eps$ and with $B^z$ near the second spin funnel.
	Here, the excited triplet state $\ket{\D^{\minus} \D^{\minus}}$ (green) hybridizes with the product state $\ket{\U^{\plus} \D^{\minus}}$ (red). The black dashed lines show the case in the absence of spin-valley coupling for comparison.
	\textbf{b}, Spin-valley-driven Rabi oscillations between $\UD$ and $\ket{\D^{\minus}\D^{\minus}}$, acquired at $B^z = 279.2$~mT, near the second spin funnel.
	We initialize and readout the state $\UD$ before and after the evolution for a variable time $t_e$ and at variable $\epsilon$.
	\textbf{c}, Control experiment corresponding to \textbf{b} where we prepare and readout the state $\ket{\D^{\plus} \U^{\minus}}$.
	No Rabi oscillations are visible. 
	\textbf{d}, Energy level diagram at large $\eps$ and with $B^z$ near the third spin funnel.
	Here, the excited triplet state $\ket{\D^{\plus} \D^{\plus}}$ (green) hybridizes with $\ket{\D^{\plus} \U^{\minus}}$ (blue).
	\textbf{e}, \textbf{f}, Same measurements as in \textbf{b} and \textbf{c}, respectively, but conducted with $B^z=380.8$~mT, near the third spin funnel.
	Here, spin-valley-driven Rabi oscillations between $\DU$ and $\ket{\D^{\plus}\D^{\plus}}$ are visible, but not between $\UD$ and $\ket{\D^{\plus}\D^{\plus}}$.
	}
	\label{fig:hotspotRabi}
\end{figure}

To investigate this possibility, we set $B^z=279.2$ mT, near the second spin funnel. 
We prepare the state $\ket{\U^{\plus} \D^{\minus}}$ in the (3,1) charge configuration by adiabatic separation of the $\ket{T_0^{\plus\minus}}$ state in the presence of a $g$-factor difference (see Methods), and then we pulse to different values of $\epsilon$ deep in (3,1).
Near $\eps=30$~mV, where $J(\eps)\approx 0$, we observe prominent Rabi oscillations with frequency $f_1=8.2$ MHz.
These oscillations occur because the excited triplet $\ket{\D^{\minus} \D^{\minus}}$ comes into resonance with the $\ket{\U^{\plus} \D^{\minus}}$ state (Figs.~\ref{fig:hotspotRabi}a-b). 
The oscillation frequency is determined by the single-spin-valley coupling in dot~1, such that $f_1=2\braket{\U^{\plus} \D^{\minus}|H^{sv}|\D^{\minus} \D^{\minus}}=2\dsv_1$, where we have set Planck's constant $h=1$.
While this is apparent within the two-electron picture, it is indeed also true in the full four-electron picture (see Methods).
When we instead prepare the state $\ket{\D^{\plus} \U^{\minus}}$ (by adiabatic separation of the $\ket{S^{\plus\minus}}$ state) at the same magnetic field, we observe no Rabi oscillations (Fig.~\ref{fig:hotspotRabi}c), which is expected since $\braket{\D^{\plus} \U^{\minus}|H^{sv}|\D^{\minus} \D^{\minus}}=0$.
Such a transition would require a pure spin flip in dot 2 and a pure valley change in dot~1.

Furthermore, when we prepare  $\ket{\D^{\plus} \U^{\minus}}$ at $B^z=380.8$~mT, near the third spin funnel, we observe prominent Rabi oscillations with $f_2=2\braket{\D^{\plus} \U^{\minus}|H^{sv}|\D^{\plus} \D^{\plus}}=2\dsv_2=15.2$ MHz, but we do not observe Rabi oscillations when we prepare $\ket{\U^{\plus} \D^{\minus}}$ (Figs.~\ref{fig:hotspotRabi}d-f). Tuning the magnetic field away from either the second or third spin funnel and performing the same experiment yield no measurable oscillations with either initial state (see Methods). We hypothesize that the features visible in Figs.~\ref{fig:hotspotRabi}c and \ref{fig:hotspotRabi}e result from spurious coupling between the states $\ket{\D^{\plus} \U^{\minus}}$ and $\ket{\U^{\plus} \D^{\minus}}$, likely due to residual exchange coupling (see Methods).

We interpret the data of Fig.~\ref{fig:hotspotRabi} as evidence for combined spin-valley Rabi oscillations between effective single-electron states.
Specifically, Fig.~\ref{fig:hotspotRabi}b corresponds to simultaneous spin flips and valley changes in dot 1, and Fig.~\ref{fig:hotspotRabi}f corresponds to simultaneous spin flips and valley changes in dot 2.
Because the spin-valley coupling is an ``always-on" matrix element, no ac magnetic or electric fields are required for these spin rotations.
Thus, spin-valley coupling offers a method for single-spin control which complements traditional methods based on oscillating fields.
In the present context, these single-spin oscillations are facilitated by singlet-triplet initialization and readout.
Note that the quantum states, which we label as effective two-electron states with the valence configuration, are actually four-electron states.
Thus, within the two-electron picture, the data of Fig.~\ref{fig:hotspotRabi} represent a single spin undergoing spin and valley oscillations.
Within the four-electron picture, however, all of the electrons in a particular dot will participate in the spin-valley oscillations.
In Fig.~\ref{fig:hotspotRabi}b, for example, all three electrons undergo collective spin-valley changes (see Methods). 
Nevertheless, the two-electron picture provides an intuitive picture of our four-electron system and accurately predicts the spin-valley couplings we observe.


\begin{figure*}[t]
	\includegraphics[width =\textwidth]{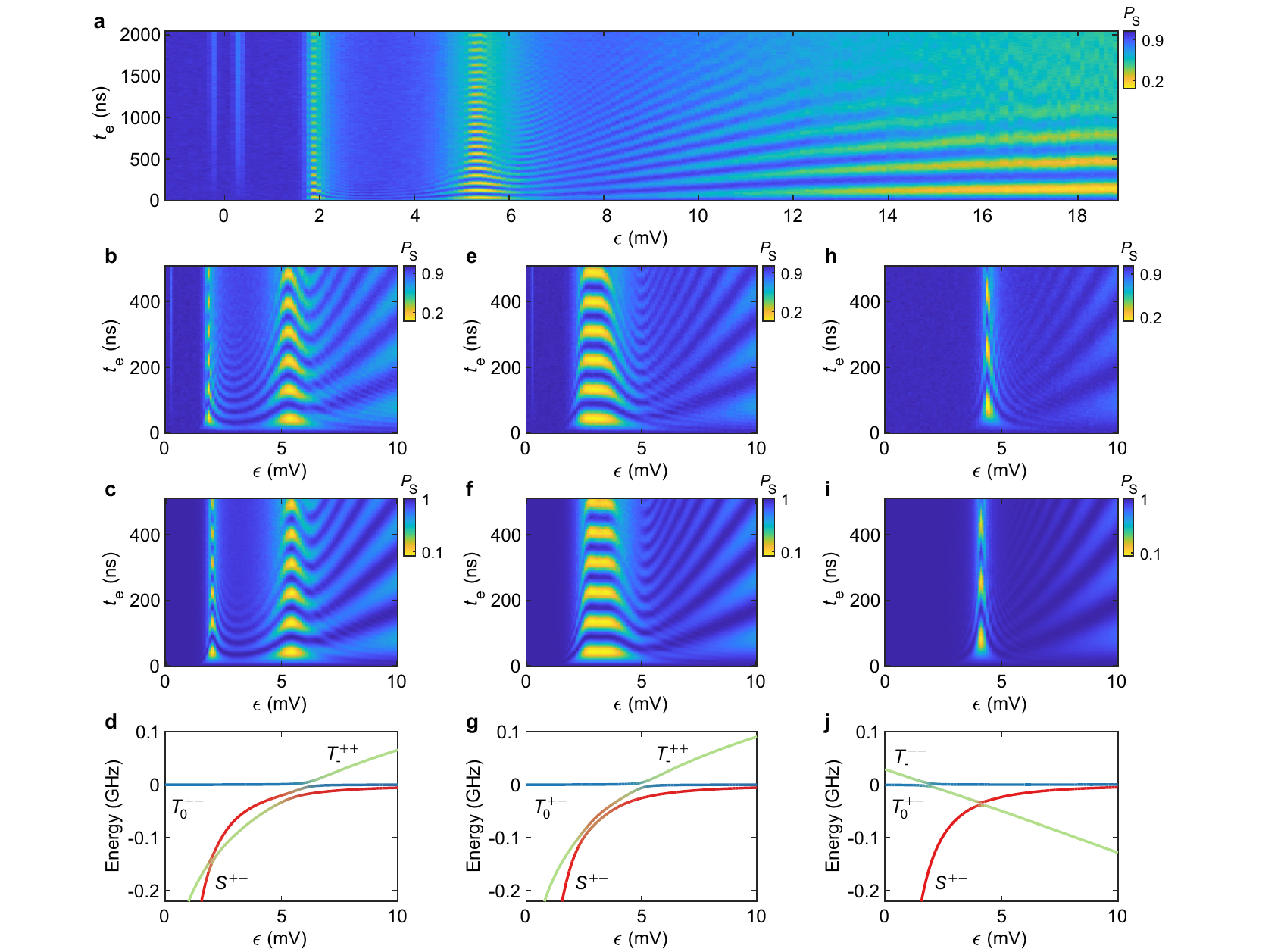}
	\caption{\stm{} oscillations driven by spin-valley coupling.
	\textbf{a}, With $B^z=372.3$ mT, near the third spin funnel, we prepare the state $\ket{S^{\plus\minus}}$ and pulse to different values of $\eps$ for different evolution times $t_e$. The first two vertical lines in the data between $\eps=-1$~mV and 1~mV result from the first and second spin funnels. The two sets of Rabi oscillations near $\eps=2$ mV and 5 mV are $S$-$T_-$ Rabi oscillations. The oscillations for $\eps>6$ mV are exchange- and gradient-driven \stz{} oscillations.
	\textbf{b}, High-resolution scan of the two sets of Rabi oscillations at $B^z=372.3$ mT.
	\textbf{c}, Numerical simulation corresponding to \textbf{b}.
	\textbf{d}, Numerically computed energies for the three relevant levels in \textbf{c}. 
	\textbf{e}, High-resolution scan of Rabi oscillations at $B^z=371.3$ mT, when the two \stm{} avoided crossings have merged.
	\textbf{f}, Numerical simulation corresponding to \textbf{e}.
	\textbf{g}, Numerically computed energies for the three relevant levels in \textbf{f}.
	\textbf{h}, High-resolution scan of Rabi oscillations at $B^z=295.0$ mT, on the second spin funnel.
	\textbf{i}, Numerical simulation corresponding to \textbf{h}.
	\textbf{j}, Numerically computed energies for the three relevant levels in \textbf{i}.
	}
	\label{fig:STmRabi}
\end{figure*}

\section{Singlet-triplet and triplet-triplet spin-valley coupling}
Given the existence of single-spin-valley matrix elements associated with dots 1 and 2 ($\dsv_1$ and $\dsv_2$, respectively), we also expect that strong singlet-triplet mixing should occur when $J(\eps)$ is appreciable, and when the singlet-triplet qubit eigenstates are effective singlets and triplets, instead of effective product states. Indeed, one can see that $\braket{S^{\plus\minus}|H^{sv}|T_-^{\minus\minus}}=\dsv_1/\sqrt{2}$, and $\braket{S^{\plus\minus}|H^{sv}|T_-^{\plus\plus}}=-\dsv_2/\sqrt{2}$ (see Methods). 

To test for this effect, we set $B^z=372.3$ mT, still near the third spin funnel. At this field, the $\ket{S^{\plus\minus}}$ and $\ket{T_-^{\plus\plus}}$ states come into resonance when $J(\eps) \gg \dez$. We initialize $\ket{S^{\plus\minus}}$ and pulse $\eps$ suddenly, which preserves the singlet state as the charge configuration changes. After variable evolution time, we pulse $\eps$ to the PSB region for readout. Against a broad background of exchange-driven and gradient-driven \stz{} oscillations, two narrow regions with pronounced and long-lived Rabi oscillations are visible (Fig.~\ref{fig:STmRabi}a).  

A high resolution scan of the two avoided crossings (Fig.~\ref{fig:STmRabi}b) agrees strikingly well with numerical simulations (see Methods) (Fig.~\ref{fig:STmRabi}c). Figure~\ref{fig:STmRabi}d shows the energies of the relevant three states in these simulations. Because $E^v_2(\Delta V_2)$ increases with $\eps$, the energy the of  $\ket{T_-^{\plus\plus}}$ state intersects that of the $\ket{S^{\plus\minus}}$ state twice at this magnetic field, yielding two different avoided crossings between these states. In either case, the Rabi frequency is $f_2^{ST}=11.3$ MHz. Note that $f_2^{ST}/f_2=0.74$, in agreement with our expectation that $\braket{S^{\plus\minus}|H^{sv}|T_-^{\plus\plus}}=-\dsv_2/\sqrt{2}$

At a slightly lower magnetic field $B^z=371.3$~mT, the two \stm{} avoided crossings merge, yielding an extended avoided crossing with pronounced Rabi oscillations of the same frequency (Fig.~\ref{fig:STmRabi}e). As above, numerical simulations (Figs.~\ref{fig:STmRabi}f-g) agree with our data. When we set $B^z=295.0$~mT, on the second spin funnel, we only observe one avoided crossing (Fig.~\ref{fig:STmRabi}h) because $E^v_1(\Delta V_1)$ decreases with $\eps$ and therefore the energy of the $\ket{T_-^{\minus\minus}}$ state only intersects that of the $\ket{S^{\plus\minus}}$ state once (Fig.~\ref{fig:STmRabi}j). The oscillation frequency $f_1^{ST}=6$ MHz and $f_1^{ST}/f_1=0.73$, again in agreement with the expectation that  $\braket{S^{\plus\minus}|H^{sv}|T_-^{\minus\minus}}=\dsv_1/\sqrt{2}$.

These data represent evidence of \stm{} Rabi oscillations, driven by spin-valley coupling. Previously, \stm{} Rabi oscillations in Si (or $S$-$T_+$ oscillations in GaAs quantum dots) have not been observed, because the most common mechanisms for \stm{} mixing occur in parameter regimes that feature extreme sensitivity to hyperfine or charge noise and significant decoherence~\cite{Stepanenko2012,Petta2010,Nichol2015}. Despite this obstacle, however, coherent oscillations of $S$-$T_-$ superposition states in Si (or $S$-$T_+$ states in GaAs) have been observed through Landau-Zener interferometry~\cite{Petta2010,Wu2014,Fogarty2018}. Our results, however, show how the spin-valley coupling overcomes the longstanding challenge of manipulating \stm{} systems and enables straightforward, universal quantum control of an $S$-$T_-$ qubit. 

\begin{figure}
\includegraphics[width =\linewidth]{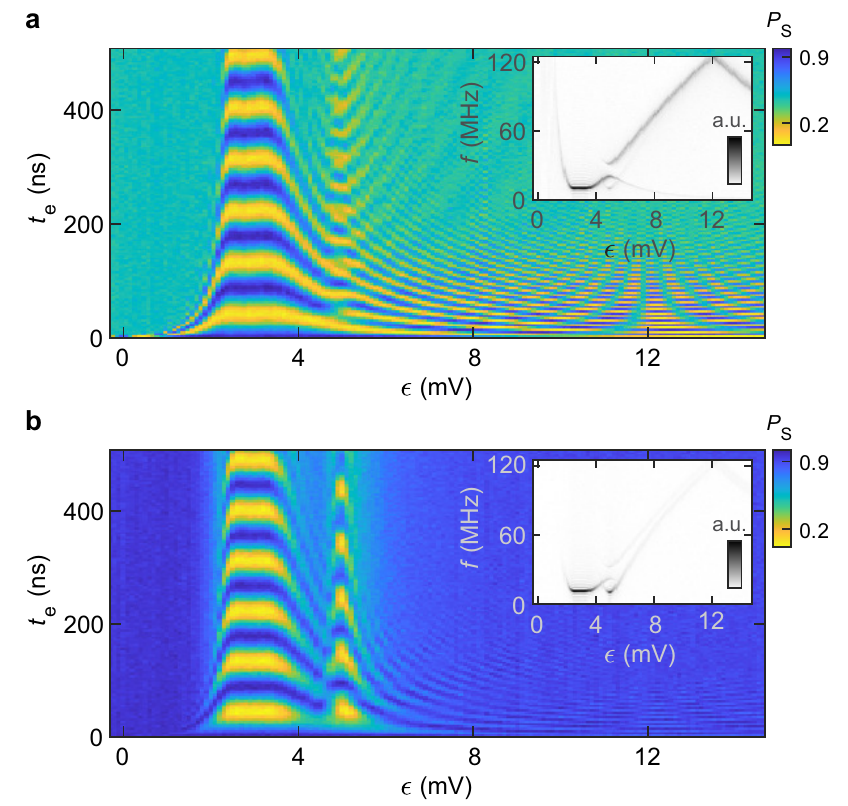}
\caption{Coherent manipulation of joint spin states through spin-valley coupling. 
        \textbf{a}, Ramsey oscillations at different values of $\eps$. Inset: Absolute value of the fast Fourier transform of the data.
        \textbf{b}, Oscillations measured with a $\pi$ pulse at $\eps=3$ mV before and after the evolution. 
        The $\pi$ pulses map $\ket{S^{\plus\minus}}$ to $\ket{T_-^{\plus\plus}}$ and vice versa. 
        The oscillations near $\eps=5$ mV correspond to $T_0$-$\tm$ Rabi oscillations, also mediated by spin-valley coupling. Inset: Absolute value of the fast Fourier transform of the data.}
\label{fig:STmOsc}
\end{figure}

To illustrate the potential of spin-valley coupling for full quantum control of $S$-$T_-$ states, we set $B^z=371.1$ mT, where the two \stm{} avoided crossings on the third spin funnel merge.
We perform a Ramsey experiment by applying a $3\pi/2$ pulse at the avoided crossing before an evolution segment of variable time and $\eps$, and a $\pi/2$ pulse after the evolution (Fig.~\ref{fig:STmOsc}a). 
Now, at values of $\eps>6$ mV, we observe relatively high-frequency oscillations corresponding to the time evolution of \stm{} superposition states.
In this range, the frequency of these oscillations increases with $\eps$ because of the $\eps$ dependence of $E^v_2$ which determines the energy of the $\ket{T_-^{\plus\plus}}$ state (Fig.~\ref{fig:STmOsc}g).
Because of the approximately linear increase of $E^v_2(\Delta V_2)$ with $\eps$, the coherence of these oscillations is nearly constant.
Such a scenario, where the coherence of an electrically-controlled multi-electron spin qubit does not decrease with frequency, is rare because such qubit splittings are usually highly non-linear functions of voltages~\cite{Petta2005,Dial2013}.
We discuss the coherence of these oscillations in further detail below.
More complicated pulse sequences are straightforward to implement.
In Methods, we discuss a spin-echo experiment, which improves the coherence time of $S$-$T_-$ superposition states by roughly a factor of 10.

A close inspection of Fig.~\ref{fig:STmOsc}a reveals an additional set of oscillations near $\eps=5$~mV. To elucidate these oscillations, we prepare the $\ket{T_-^{\plus\plus}}$ state by applying a $\pi$ pulse at the $S$-$T_-$ avoided crossing before pulsing to different values of $\epsilon$ (Fig.~\ref{fig:STmOsc}b). After a variable evolution time, we apply another $\pi$ pulse on resonance, effectively mapping $\ket{T_-^{\plus\plus}}$ back to $\ket{S^{\plus\minus}}$. The data from this experiment reveal another avoided crossing near $\eps=5$~mV, which we attribute to mixing between $\ket{T_-^{\plus\plus}}$ and $\ket{T_0^{\plus\minus}}$ (Fig.~\ref{fig:STmOsc}g). These triplet-triplet oscillations have a similar frequency as the main avoided crossing, because $\braket{T_0^{\plus\minus}|H^{sv}|T_-^{\plus\plus}}=\dsv_2/\sqrt{2}$ (and $\braket{T_0^{\plus\minus}|H^{sv}|T_-^{\minus\minus }}=\dsv_1/\sqrt{2}$). Such triplet-triplet oscillations represent a previously-unexplored spin-state transition driven by spin-valley coupling between effective joint spin states in a double quantum dot. 

As a final confirmation that these effects originate from spin-orbit coupling, we measure the dependence of $\dsv_1$ and $\dsv_2$ on the direction of the external in-plane magnetic field (see Methods). We find that the magnitudes of both spin-valley couplings oscillate with the in-plane field angle as expected for spin-orbit coupling (see Methods).
In addition to corroborating the notion that the spin-valley coupling in our system originates from spin-orbit coupling, these data reveal differences in the relative strengths of the Rashba and Dresselhaus spin-orbit couplings and valley phase differences between quantum dots~\cite{zhang2020anisotropy}. 

\begin{figure*}[t]
\includegraphics[width =\textwidth]{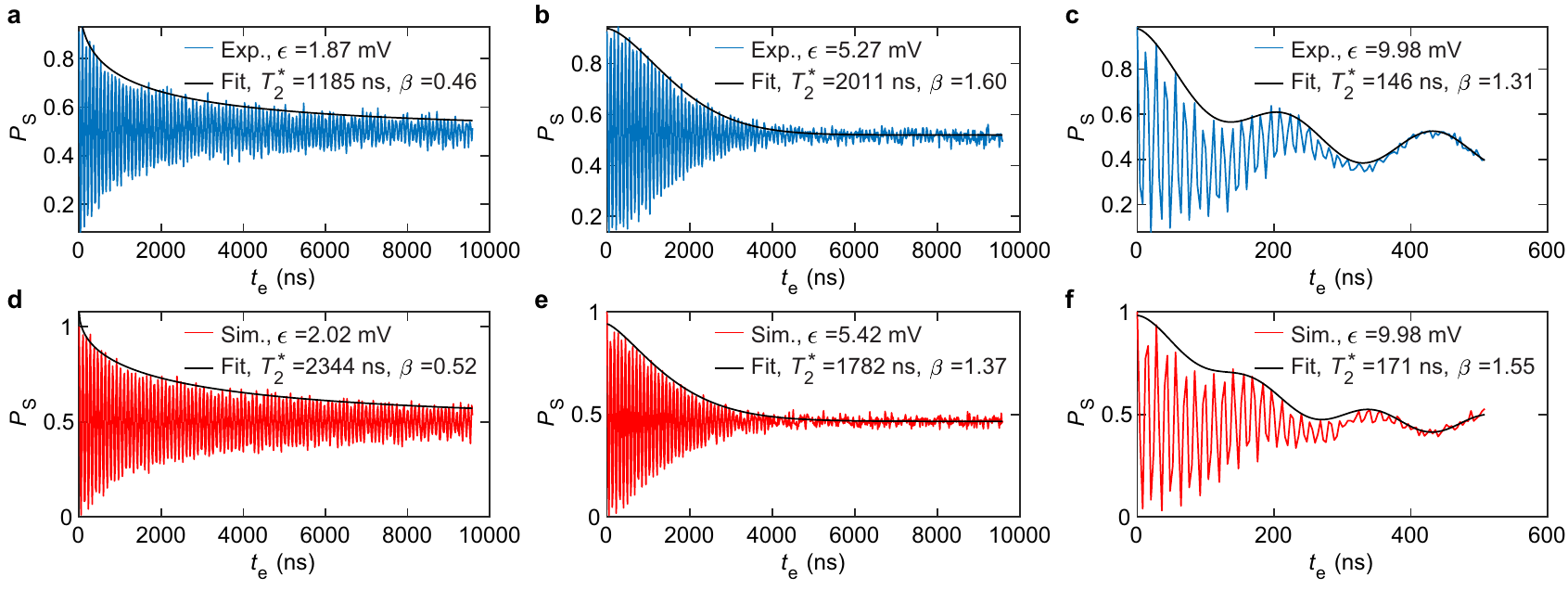}
\caption{Measured and simulated \stm{} oscillations on the third spin funnel, where the two avoided crossings are separate.
        \textbf{a}, Measured Rabi oscillations at the left avoided crossing. The data are fit to an equation of the form $P_S(t)=P_0+A \cos(\omega t +\phi) \exp\left[-(t/T_2^*)^\beta\right]$, with $P_0$, $A$, $\omega$, $\phi$, $T_2^*$, and $\beta$ as fit parameters. The overlaid black line is $P_0+A  \exp\left[-(t/T_2^*)^\beta\right]$. 
        \textbf{b}, Measured Rabi oscillations at the right avoided crossing.
        \textbf{c}, Measured Ramsey oscillations near $\epsilon=10$ mV. The slow background oscillations are \stz{} oscillations. We fit the data to an equation of the form $P_S(t)=P_0+A \cos(\omega t +\phi) \exp\left[-(t/T_2^*)^\beta\right]+A_2 \cos(\omega_2 t + \phi_2)$, where $A_2$, $\omega_2$, and $\phi_2$ describe the \stz{} oscillations. The black line is $P_0+A \exp\left[-(t/T_2^*)^\beta\right]+A_2 \cos(\omega_2 t + \phi)$.
        \textbf{d}, Numerical simulations corresponding to panel \textbf{a} and fit.
        We choose the value of $\eps$ corresponding to the local minimum in the oscillation frequency in the region near $\eps=2$~mV, corresponding to the location of the left avoided crossing. 
        \textbf{e}, Numerical simulation corresponding to \textbf{b} and fit.
        We choose the value of $\eps$ corresponding to the local minimum in the oscillation frequency in the region near $\eps=5$~mV, corresponding to the location of the right avoided crossing.
        \textbf{f}, Numerical simulation corresponding to \textbf{c}. Here we chose the same value of $\epsilon$ as in \textbf{c}. 
        In all cases, the measured and simulated values of $T_2^*$ and $\beta$ show reasonably good agreement with each other.}
\label{fig:coherence}
\end{figure*}

\section{Spin-valley oscillation coherence}
Finally, we analyze the coherence of the observed spin-valley \stm{} oscillations. We set $B^z=370.4$ mT, on the third spin funnel, where the two avoided crossings remain separate, and perform Ramsey experiments using a similar procedure to that described above. Figures~\ref{fig:coherence}a-c show the time series obtained from these experiments at three specific detunings: the two values of $\eps$ where the \stm{} avoided crossings occur, and one value of $\eps$ deep in (3,1). Both avoided crossings feature long-lived coherence (Fig.~\ref{fig:coherence}a-b), with Rabi oscillations persisting for at least several microseconds. 

The decay of Rabi oscillations at both avoided crossings is clearly non-Gaussian  (Figs.~\ref{fig:coherence}a-b), especially at the left avoided crossing. We attribute this fact to the non-Gaussian nature of fluctuations in the energy splitting at the avoided crossing~\cite{Mahklin2004}. For the left avoided crossing, we expect this effect to be pronounced, because the energy splitting in the vicinity of the avoided crossing depends strongly and non-linearly on $\eps$. In both cases, however, our numerical simulations (Figs.~\ref{fig:coherence}d-e) (see Methods) reproduce both the observed coherence times and decay shapes. The simulations are based on solving the Schr\"odinger equation for the time evolution of the relevant three levels ($\ket{S^{\plus\minus}}$, $\ket{T_0^{\plus\minus}}$, and either $\ket{T_-^{\minus\minus}}$ or $\ket{T_-^{\plus\plus}}$) and include independent, random, quasistatic magnetic-field fluctuations for each electron (with rms value $\sigma_B=8$ $\mu$T, corresponding to an inhomogeneuosly broadened single-spin coherence time of 1 $\mu$s), as well as independent gate-voltage fluctuations on each plunger gate (with rms values 34 $\mu$V and 92 $\mu$V for $P_1$ and $P_2 $, respectively, see Methods). 

At $\epsilon=10$ mV, where the oscillations reflect the time evolution of \stm{} superposition states (in addition to the slower, \stz{} dynamics), the decay occurs over a timescale of about a hundred nanoseconds (Fig.~\ref{fig:coherence}c). Because we assume that the valley splitting $E^v_2(\Delta V_2)$ depends only on $V_2$, we attribute this decay to electrical noise associated with $V_2$. 
Because of this, and because a similar data set on the second spin funnel reflects the noise associated with $V_1$, we can determine the effective gate voltage fluctuations separately for each dot. As above, our numerical simulations reproduce the approximate decay time and shape (Figs.~\ref{fig:coherence}f). In the future, this ability to independently resolve electrical fluctuations on separate quantum dots will complement measurements of charge noise based on exchange coupling~\cite{Dial2013,Connors2021charge,Jock2021Valley}, which depends on the difference between chemical potentials of two dots. 

We have performed similar analyses for the second spin funnel and for the third spin funnel at a magnetic field where the avoided crossings have merged (see Methods). In all cases, we achieve roughly the same level of agreement between experiment and simulation, suggesting that our simple three-level model with independent quasi-static electrical and magnetic noise associated with each dot successfully explains the features we observe.

\section{Conclusion}
We have demonstrated that spin-valley coupling, a phenomenon usually considered a hindrance in Si spin-qubit systems, can enable coherent manipulation of both single- and multi-spin states in Si quantum dots.
Our measurements reveal a simple method to control effective single-spin states without ac magnetic or electric fields, as well as \stm{} Rabi and Ramsey oscillations and previously-unexplored triplet-triplet oscillations.
Because of its simplicity, this method of coherent spin manipulation can potentially supplement existing control methods in many existing qubit encodings.
Coherent spin manipulation via spin-valley coupling can also immediately add an important capability to charge-noise spectroscopy in Si quantum dots by offering a method to explore local charge noise in quantum-dot spin qubits, complementing standard techniques based on exchange coupling.
Finally, our measurements present a striking confirmation of how simple Hubbard models and effective valence-electron pictures can apply to multi-electron systems undergoing complex, collective dynamics. 

More broadly, our measurements introduce a new family of spin qubits, including effective single-spin, singlet-triplet, and triplet-triplet qubits based on spin-valley coupling.
The long coherence times and ease of control associated with these qubits warrants future work to fully characterize their properties, including gate fidelities. 
The implementation of systems with multiple spin-valley qubits will likely require that the different spin qubits either have the same valley splittings or different Zeeman energies, which could be potentially achieved through heterostructure engineering~\cite{sasaki2009well,Chen2021,mcjunkin2021valley,neyens2018critical}, voltage control of valley splittings~\cite{Yang2013,Chen2021,hollmann_large_2020}, or local magnetic field control~\cite{Borjans2020,astner2017coherent}.
Altogether, these results highlight the suprising usefulness of the valley degree of freedom, a unique feature of electrons in Si, for quantum information processing.

\section{Data Availability}
The data that support the findings of this study are available from the corresponding author upon reasonable request.

\bibliography{STm_bib.bib}

\section{Acknowledgments}
We thank Lisa F. Edge of HRL Laboratories, LLC. for the epitaxial growth of the SiGe material.
This work was sponsored the Army Research Office under grants W911NF-16-1-0260 and W911NF-19-1-0167. The views and conclusions contained in this document are those of the authors and should not be interpreted as representing the official policies, either expressed or implied, of the Army Research Office or the U.S. Government. The U.S. Government is authorized to reproduce and distribute reprints for Government purposes notwithstanding any copyright notation herein.

\section{Author Contributions}
X.C., E.J.C., and J.M.N. conceptualized the experiment, conducted the investigation, and participated in writing. J.M.N. supervised the effort.

\end{document}


\title{Methods}
	
\author{Xinxin Cai}
\thanks{These authors contributed equally.}
\affiliation{Department of Physics and Astronomy, University of Rochester, Rochester, NY, 14627 USA}
	
\author{Elliot J. Connors}
\thanks{These authors contributed equally.}
\affiliation{Department of Physics and Astronomy, University of Rochester, Rochester, NY, 14627 USA}
	
	
\author{John M. Nichol}
\email{john.nichol@ur.rochester.edu}
\affiliation{Department of Physics and Astronomy, University of Rochester, Rochester, NY, 14627 USA}
	
	
\pacs{}
	
	
\onecolumngrid
	

\section{Methods}

\subsection{Four-electron states}
Here we explicitly list the four-electron states involved in this work and compute the spin-valley matrix elements between them. We assume that the four-electron states can be represented as Slater determinants, or as linear combinations of Slater determinants, where only the lowest two valley levels in each dot may be occupied.

Within this approximation, the wavefunction for the $\ket{\tm^{\minus\minus}}$ state is given by the following determinant:
\begin{equation}
\begin{split}
\ket{\tm^{\minus\minus}}&\equiv\ket{\U^{1\plus}\D^{1\plus}\D^{1\minus}\D^{2\minus}} = \frac{1}{\sqrt{4!}}
\begin{vmatrix} 
\U_1^{1\plus} & \D_1^{1\plus} & \D_1^{1\minus} & \D_1^{2\minus}  \\ 
\U_2^{1\plus} & \D_2^{1\plus} & \D_2^{1\minus} & \D_2^{2\minus}  \\ 
\U_3^{1\plus} & \D_3^{1\plus} & \D_3^{1\minus} & \D_3^{2\minus}  \\ 
\U_4^{1\plus} & \D_4^{1\plus} & \D_4^{1\minus} & \D_4^{2\minus}  \\ 
\end{vmatrix}\\
&=\frac{1}{\sqrt{4!}}(|\U^{1\plus}_1\D^{2\minus}_2\D^{1\plus}_3\D^{1\minus}_4\rangle-|\U^{1\plus}_1\D^{2\minus}_2\D^{1\minus}_3\D^{1\plus}_4\rangle-|\U^{1\plus}_1\D^{1\plus}_2\D^{2\minus}_3\D^{1\minus}_4\rangle+|\U^{1\plus}_1\D^{1\minus}_2\D^{2\minus}_3\D^{1\plus}_4\rangle+\\ 
&|\U^{1\plus}_1\D^{1\plus}_2\D^{1\minus}_3\D^{2\minus}_4\rangle-|\U^{1\plus}_1\D^{1\minus}_2\D^{1\plus}_3\D^{2\minus}_4\rangle-|\D^{2\minus}_1\U^{1\plus}_2\D^{1\plus}_3\D^{1\minus}_4\rangle+|\D^{2\minus}_1\U^{1\plus}_2\D^{1\minus}_3\D^{1\plus}_4\rangle+\\ 
&|\D^{1\plus}_1\U^{1\plus}_2\D^{2\minus}_3\D^{1\minus}_4\rangle-|\D^{1\minus}_1\U^{1\plus}_2\D^{2\minus}_3\D^{1\plus}_4\rangle-|\D^{1\plus}_1\U^{1\plus}_2\D^{1\minus}_3\D^{2\minus}_4\rangle+|\D^{1\minus}_1\U^{1\plus}_2\D^{1\plus}_3\D^{2\minus}_4\rangle+\\ 
&|\D^{2\minus}_1\D^{1\plus}_2\U^{1\plus}_3\D^{1\minus}_4\rangle-|\D^{2\minus}_1\D^{1\minus}_2\U^{1\plus}_3\D^{1\plus}_4\rangle-|\D^{1\plus}_1\D^{2\minus}_2\U^{1\plus}_3\D^{1\minus}_4\rangle+|\D^{1\minus}_1\D^{2\minus}_2\U^{1\plus}_3\D^{1\plus}_4\rangle+\\ 
&|\D^{1\plus}_1\D^{1\minus}_2\U^{1\plus}_3\D^{2\minus}_4\rangle-|\D^{1\minus}_1\D^{1\plus}_2\U^{1\plus}_3\D^{2\minus}_4\rangle-|\D^{2\minus}_1\D^{1\plus}_2\D^{1\minus}_3\U^{1\plus}_4\rangle+|\D^{2\minus}_1\D^{1\minus}_2\D^{1\plus}_3\U^{1\plus}_4\rangle+\\ 
&|\D^{1\plus}_1\D^{2\minus}_2\D^{1\minus}_3\U^{1\plus}_4\rangle-|\D^{1\minus}_1\D^{2\minus}_2\D^{1\plus}_3\U^{1\plus}_4\rangle-|\D^{1\plus}_1\D^{1\minus}_2\D^{2\minus}_3\U^{1\plus}_4\rangle+|\D^{1\minus}_1\D^{1\plus}_2\D^{2\minus}_3\U^{1\plus}_4\rangle).
\end{split}
\end{equation}
In the determinant, the subscript indicates the electron number, and the superscript indicates the valley state. $^{1\minus}$ represents the lower  valley of dot 1, $^{1\plus}$ is the higher valley of dot 1, $^{2\minus}$ is the lower valley of dot 2, and $^{2\plus}$ is the higher valley of dot 2.

Similarly, 
\begin{equation}
\begin{split}
\ket{\tm^{\plus\plus}}&\equiv\ket{\U^{1\minus}\D^{1\minus}\D^{1\plus}\D^{2\plus}}
=\frac{1}{\sqrt{4!}}
\begin{vmatrix} 
\U_1^{1\minus} & \D_1^{1\minus} & \D_1^{1\plus} & \D_1^{2\plus}  \\ 
\U_2^{1\minus} & \D_2^{1\minus} & \D_2^{1\plus} & \D_2^{2\plus}  \\ 
\U_3^{1\minus} & \D_3^{1\minus} & \D_3^{1\plus} & \D_3^{2\plus}  \\ 
\U_4^{1\minus} & \D_4^{1\minus} & \D_4^{1\plus} & \D_4^{2\plus}  \\ 
\end{vmatrix}
\end{split}
\end{equation}

\begin{equation}
\begin{split}
\ket{\tm^{\plus\minus}}&\equiv\ket{\U^{1\minus}\D^{1\minus}\D^{1\plus}\D^{2\minus}}
=\frac{1}{\sqrt{4!}}
\begin{vmatrix} 
\U_1^{1\minus} & \D_1^{1\minus} & \D_1^{1\plus} & \D_1^{2\minus}  \\ 
\U_2^{1\minus} & \D_2^{1\minus} & \D_2^{1\plus} & \D_2^{2\minus}  \\ 
\U_3^{1\minus} & \D_3^{1\minus} & \D_3^{1\plus} & \D_3^{2\minus}  \\ 
\U_4^{1\minus} & \D_4^{1\minus} & \D_4^{1\plus} & \D_4^{2\minus}  \\ 
\end{vmatrix}
\end{split}
\end{equation}

\begin{equation}
\begin{split}
\UD&\equiv\ket{\U^{1\minus}\D^{1\minus}\U^{1\plus}\D^{2\minus}}=\frac{1}{\sqrt{4!}}
\begin{vmatrix} 
\U_1^{1\minus} & \D_1^{1\minus} & \U_1^{1\plus} & \D_1^{2\minus}  \\ 
\U_2^{1\minus} & \D_2^{1\minus} & \U_2^{1\plus} & \D_2^{2\minus}  \\ 
\U_3^{1\minus} & \D_3^{1\minus} & \U_3^{1\plus} & \D_3^{2\minus}  \\ 
\U_4^{1\minus} & \D_4^{1\minus} & \U_4^{1\plus} & \D_4^{2\minus}  \\ 
\end{vmatrix}
\end{split}
\end{equation}
and 

\begin{equation}
\begin{split}
\DU&\equiv\ket{\U^{1\minus}\D^{1\minus}\D^{1\plus}\U^{2\minus}}
=\frac{1}{\sqrt{4!}}
\begin{vmatrix} 
\U_1^{1\minus} & \D_1^{1\minus} & \D_1^{1\plus} & \U_1^{2\minus}  \\ 
\U_2^{1\minus} & \D_2^{1\minus} & \D_2^{1\plus} & \U_2^{2\minus}  \\ 
\U_3^{1\minus} & \D_3^{1\minus} & \D_3^{1\plus} & \U_3^{2\minus}  \\ 
\U_4^{1\minus} & \D_4^{1\minus} & \D_4^{1\plus} & \U_4^{2\minus}  \\ 
\end{vmatrix}
\end{split}.
\end{equation}
Finally, to construct the singlet state, we take a linear combination of Slater determinants $\ket{S^{\plus\minus}}=\frac{1}{\sqrt{2}} \left(\UD-\DU\right)$.
Likewise, we construct $\ket{T_0^{\plus\minus}}=\frac{1}{\sqrt{2}} \left(\UD+\DU\right)$.

We suppose that spin-orbit coupling creates the following single-spin-valley matrix elements:
\begin{equation}
\begin{split}
\label{equ:dsv}
\braket{\cdots \U_i^{1\minus} \cdots |H^{sv}|\cdots \D_i^{1\plus} \cdots}=\dsv_1,\\
\braket{\cdots \U_i^{2\minus} \cdots |H^{sv}|\cdots \D_i^{2\plus} \cdots}=\dsv_2.
\end{split}
\end{equation}
It is straightforward to show that $\braket{\U^{\plus} \D^{\minus}|H^{sv}|\tm^{\minus\minus}} = \dsv_1$ and $\braket{\D^{\plus} \U^{\minus}|H^{sv}|\tm^{\plus\plus}} = \dsv_2$. One can also see that $\braket{\D^{\plus} \U^{\minus}|H^{sv}|\tm^{\minus\minus}} = \braket{\U^{\plus} \D^{\minus}|H^{sv}|\tm^{\plus\plus}}=0$. To form the matrix elements with the singlet state, we take $\braket{S^{\plus\minus}|H^{sv}|\tm^{\minus\minus}} = \frac{1}{\sqrt{2}} (\braket{\U^{\plus} \D^{\minus}|H^{sv}|\tm^{\minus\minus}} - \braket{\D^{\plus} \U^{\minus}|H^{sv}|\tm^{\minus\minus}})=\frac{\dsv_1}{\sqrt{2}}$, and $\braket{S^{\plus\minus}|H^{sv}|\tm^{\plus\plus}} = \frac{1}{\sqrt{2}} (\braket{\U^{\plus} \D^{\minus}|H^{sv}|\tm^{\plus\plus}} - \braket{\D^{\plus} \U^{\minus}|H^{sv}|\tm^{\plus\plus}})=-\frac{\dsv_2}{\sqrt{2}}$. Also, one has $\braket{T_0^{\plus\minus}|H^{sv}|\tm^{\minus\minus}} = \frac{\dsv_1}{\sqrt{2}}$ and $\braket{T_0^{\plus\minus}|H^{sv}|\tm^{\plus\plus}} = \frac{\dsv_2}{\sqrt{2}}$. Thus, the simple two-electron effective wavefunctions listed in the main text behave similarly to the full four-electron wavefunctions within the approximation that only the lowest valley levels are involved in the Slater determinants. 

\subsection{Hamiltonian}
We may describe our system of four electrons in two dots with the following Hamiltonian $H=H^{Hub}+H^s+H^v+H^{sv}$.
\begin{align}
H^{Hub}=&\sum_{i=1}^{2}\left[ \frac{\tilde{U}}{2}n_i(n_i-1)+\mu_i n_i \right] 
+U_C n_1 n_2 +
\sum_{\sigma=\U \D} \sum_{\tau_{1,2}=+,-} t (c_{1,\sigma,\tau_1}^{\dagger}c_{2,\sigma,\tau_2}+\textnormal{h.c.}), \nonumber\\
H^v=&\sum_{i=1}^2\sum_{\sigma=\U,\D} (c_{i,\sigma,+}^{\dagger}c_{i,\sigma,+}) E^v_i,\\
H^s=&\frac{1}{2}\sum_{i=1}^2\sum_{\tau=+,-} (c_{i,\U,\tau}^{\dagger}c_{i,\U,\tau}-c_{i,\D,\tau}^{\dagger}c_{i,\D,\tau}) g_i \mu_B B^z. \nonumber
\end{align}
Here $c_{i,\sigma}^{\dagger}$ is a fermionic operater that creates an electron in dot $i$ with spin $\sigma$, and $n_i=\sum_{\sigma}c_{i,\sigma}c_{i,\sigma}^{\dagger}$ is the number operator for dot $i$. $\mu_i$ is the gate-controlled energy of dot $i$, $\tilde{U}$ is the on-site Coulomb energy, $U_C$ is the nearest neighbor Coulomb energy, and $t$ is the hopping energy between dots. Here, $\sigma=\U,\D$ signifies the spin state of the electron, and $\tau=+,-$ signifies the valley state. Tunneling conserves spin but not necessarily the valley state. 

We first write the matrix form of $H'=H^{Hub}+H^s+H^v$ in the following basis,
\begin{align}
\ket{S}\equiv\ket{\U^{1-} \D^{1-} \U^{1+} \D^{1+}} = (c_{1,\U,-}^{\dagger}c_{1,\D,-}^{\dagger}c_{1,\U,+}^{\dagger}c_{1,\D,+}^{\dagger}) \ket{0} \nonumber\\
\ket{\U^+ \D^-}\equiv\ket{\U^{1-} \D^{1-} \U^{1+} \D^{2-}} = (c_{1,\U,-}^{\dagger}c_{1,\D,-}^{\dagger}c_{1,\U,+}^{\dagger}c_{2,\D,-}^{\dagger}) \ket{0} \nonumber \\
\ket{\D^+ \U^-}\equiv\ket{\U^{1-} \D^{1-} \D^{1+} \U^{2-}} = (c_{1,\U,-}^{\dagger}c_{1,\D,-}^{\dagger}c_{1,\D,+}^{\dagger}c_{2,\U,-}^{\dagger}) \ket{0} \nonumber \\
\ket{\D^+ \D^-}\equiv\ket{\U^{1-} \D^{1-} \D^{1+} \D^{2-}} = (c_{1,\U,-}^{\dagger}c_{1,\D,-}^{\dagger}c_{1,\D,+}^{\dagger}c_{2,\D,-}^{\dagger}) \ket{0} \\
\ket{\D^- \D^-}\equiv\ket{\U^{1+} \D^{1+} \D^{1-} \D^{2-}} = (c_{1,\U,+}^{\dagger}c_{1,\D,+}^{\dagger}c_{1,\D,-}^{\dagger}c_{2,\D,-}^{\dagger}) \ket{0} \nonumber\\
\ket{\D^+ \D^+}\equiv\ket{\U^{1-} \D^{1-} \D^{1+} \D^{2+}} = (c_{1,\U,-}^{\dagger}c_{1,\D,-}^{\dagger}c_{1,\D,+}^{\dagger}c_{2,\D,+}^{\dagger}) \ket{0} \nonumber \\
\ket{\D^- \D^+}\equiv\ket{\U^{1+} \D^{1+} \D^{1-} \D^{2+}} = (c_{1,\U,+}^{\dagger}c_{1,\D,+}^{\dagger}c_{1,\D,-}^{\dagger}c_{2,\D,+}^{\dagger}) \ket{0} \nonumber
\end{align}
Here $\ket{0}$ is the state with no electrons. Since the $c$ operators are fermionic operators, each of these states is a Slater determinant as described above. In this basis,
\begin{align}
H'=\begin{pmatrix}
a+E^v_1 & t & -t & 0 & 0 & 0 & 0\\
t & b+\frac{\dez}{2} & 0 & 0 & 0 & 0 & 0\\
-t & 0 & b-\frac{\dez}{2}  & 0 & 0 & 0 & 0\\
0  & 0 & 0 & b-\bar{E}_Z & 0 & 0 & 0\\
0  & 0 & 0 & 0 & b+E^v_1-\bar{E}_Z & 0 & 0 \\
0  & 0 & 0 & 0 & 0 & b+E^v_2-\bar{E}_Z & 0\\
0  & 0 & 0 & 0 & 0 & 0 & b+E^v_1+E^v_2-\bar{E}_Z
\end{pmatrix},  
\end{align}
where $a=6\tilde{U}+ 4\mu_1+E^v_1$,  $b=3\tilde{U}+3\mu_1+\mu_2+3U_C+E^v_1$, $\dez=g_1 \mu_B B^z - g_2 \mu_B B^z$, and $\bar{E}_Z=\frac{1}{2}\left(g_1 + g_2\right) \mu_B B^z$. Next, subtracting the constant energy shift $b$, transforming to the basis spanned by $\{\ket{S}$, $\ket{S^{+-}}$, $\ket{T_0^{+-}}$, $\ket{\tm^{+-}}$, $\ket{\tm^{--}}$, $\ket{\tm^{++}}$, $\ket{\tm^{-+}}\}$, and setting $\epsilon=3(\tilde{U}-U_C)+\mu_1-\mu_2+E^v_1$ (in this section, $\epsilon$ is in units of energy), we have
\begin{align}
H'=\begin{pmatrix}
\epsilon & \sqrt{2}t & 0 & 0 & 0 & 0 & 0\\
\sqrt{2} t & 0 & \frac{\dez}{2} & 0 & 0 & 0 & 0\\
0 & \frac{\dez}{2}  & 0 & 0 & 0 & 0 & 0\\
0  & 0 & 0 & -\bar{E}_Z & 0 & 0 & 0\\
0  & 0 & 0 & 0 & E^v_1-\bar{E}_Z & 0 & 0 \\
0  & 0 & 0 & 0 & 0 & E^v_2-\bar{E}_Z & 0\\
0  & 0 & 0 & 0 & 0 & 0 & E^v_1+E^v_2-\bar{E}_Z
\end{pmatrix},
\end{align}
which is a conventional singlet-triplet qubit Hamiltonian in the sector spanned by the first three states. To see this, let us now change the basis of the singlet states to the adiabatic basis, $\ket{S_+}=\cos\theta\ket{S}+\sin\theta\ket{S^{+-}}$ and $\ket{S_-}=-\sin\theta\ket{S}+\cos\theta\ket{S^{+-}}$, where $\cos2\theta=\frac{\epsilon}{\sqrt{\epsilon^2+8t^2}}$, and $\sin2\theta=\frac{2\sqrt{2}t}{\sqrt{\epsilon^2+8t^2}}$. In the basis $\{\ket{S_+}$, $\ket{S_-}$, $\ket{T_0^{+-}}$, $\ket{\tm^{+-}}$, $\ket{\tm^{--}}$, $\ket{\tm^{++}}$, $\ket{\tm^{-+}}\}$, 
\begin{align}
H'= 
\begin{pmatrix}
E_+(\epsilon) & 0 & \frac{\dez}{2}\sin\theta & 0 & 0 & 0 & 0\\
0 & E_-(\epsilon) & \frac{\dez}{2}\cos\theta & 0 & 0 & 0 & 0\\
\frac{\dez}{2}\sin\theta & \frac{\dez}{2} \cos\theta  & 0 & 0 & 0 & 0 & 0\\
0  & 0 & 0 & -\bar{E}_Z & 0 & 0 & 0\\
0  & 0 & 0 & 0 & E^v_1-\bar{E}_Z & 0 & 0 \\
0  & 0 & 0 & 0 & 0 & E^v_2-\bar{E}_Z & 0\\
0  & 0 & 0 & 0 & 0 & 0 & E^v_1+E^v_2-\bar{E}_Z
\end{pmatrix},
\end{align}
where $E_{\pm}(\epsilon)=\frac{1}{2}\left(\epsilon \pm \sqrt{\epsilon^2+8t^2}\right)$. The exchange coupling $J(\epsilon)= -E_-(\epsilon)$ and $E_+(\epsilon) = \epsilon + J(\epsilon)$. Assuming that $\epsilon \gg t$, $\cos(\theta)\approx 1$, and we may approximate $\ket{S_-}\approx \ket{S^{+-}}$. We may also neglect the excited singlet state, leaving the final Hamiltonian in the basis  $\{\ket{S^{+-}}$, $\ket{T_0^{+-}}$, $\ket{\tm^{+-}}$, $\ket{\tm^{--}}$, $\ket{\tm^{++}}$, $\ket{\tm^{-+}}\}$
\begin{align}
H'=
\begin{pmatrix}
E_-(\epsilon) & \frac{\dez}{2} & 0 & 0 & 0 & 0\\
\frac{\dez}{2}  & 0 & 0 & 0 & 0 & 0\\
0 & 0 & -\bar{E}_Z & 0 & 0 & 0\\
0 & 0 & 0 & E^v_1-\bar{E}_Z & 0 & 0 \\
0 & 0 & 0 & 0 & E^v_2-\bar{E}_Z & 0\\
0 & 0 & 0 & 0 & 0 & E^v_1+E^v_2-\bar{E}_Z
\end{pmatrix}.
\end{align}

The spin-valley matrix elements are those obtained above. The total Hamiltonian $H=H'+H^{sv}$ is 
\begin{align}
H=
\begin{pmatrix}
E_-(\epsilon) & \frac{\dez}{2} & 0 &  \frac{\dsv_1}{\sqrt{2}} &  -\frac{\dsv_2}{\sqrt{2}} & 0\\
\frac{\dez}{2}  & 0 & 0 & \frac{\dsv_1}{\sqrt{2}} & \frac{\dsv_2}{\sqrt{2}} & 0\\
0 & 0 & -\bar{E}_Z & 0 & 0 & 0\\
\frac{(\dsv_1)^*}{\sqrt{2}} & \frac{(\dsv_1)^*}{\sqrt{2}} & 0 & E^v_1-\bar{E}_Z & 0 & 0 \\
-\frac{(\dsv_2)^*}{\sqrt{2}} & \frac{(\dsv_2)^*}{\sqrt{2}} & 0 & 0 & E^v_2-\bar{E}_Z & 0\\
0 & 0 & 0 & 0 & 0 & E^v_1+E^v_2-\bar{E}_Z
\end{pmatrix}.
\end{align}

In the above analysis, the exchange coupling depends on not just the detuning but also $\Ev_1$. Both quantities depend on gate voltages. To determine the relative size of the fluctuations in the exchange coupling due to the effects of electrical noise on both quantities, we define $\epsilon'\equiv 3(\tilde{U}-U_C)+ \mu_1-\mu_2$, which is the experimentally controlled detuning.
Now $E_-(\epsilon) = E_-(\epsilon'+E^v_1(\Delta V_1))$. To assess the impact of the additional voltage dependence in $E_v^1$ on the exchange coupling, we note 
\begin{align}
\beta_{\eps}=\frac{dJ}{d (\epsilon'+E^v_1)}\frac{d\epsilon'}{d \Delta V_1}=\frac{1}{2}\left(1-\frac{\epsilon'+E^v_1}{\sqrt{(\epsilon'+E^v_1)^2+8t^2}}\right)\alpha_1,  
\end{align}
where $\alpha_1$ is the lever arm of dot 1. Also,
\begin{align}
\beta_{v}=\frac{dJ}{d (\epsilon'+E^v_1)}\frac{dE^v_1}{d \Delta V_1}=\frac{1}{2}\left(\frac{\epsilon'+E^v_1}{\sqrt{(\epsilon'+E^v_1)^2+8t^2}}-1\right)  \frac{dE^v_1}{d \Delta V_1}.
\end{align}
Thus, $\left|\beta_{\eps}/\beta_v\right|=\alpha_1/ (dE^v_1/d\Delta V_1) > 200$. The additional dependence of the exchange coupling $J$ on voltages through the valley splitting can safely be ignored in our simulations.

\subsection{Spin funnel measurement and analysis}
\label{sec:SF}

\begin{figure}[ht!]
	\includegraphics[width =\textwidth]{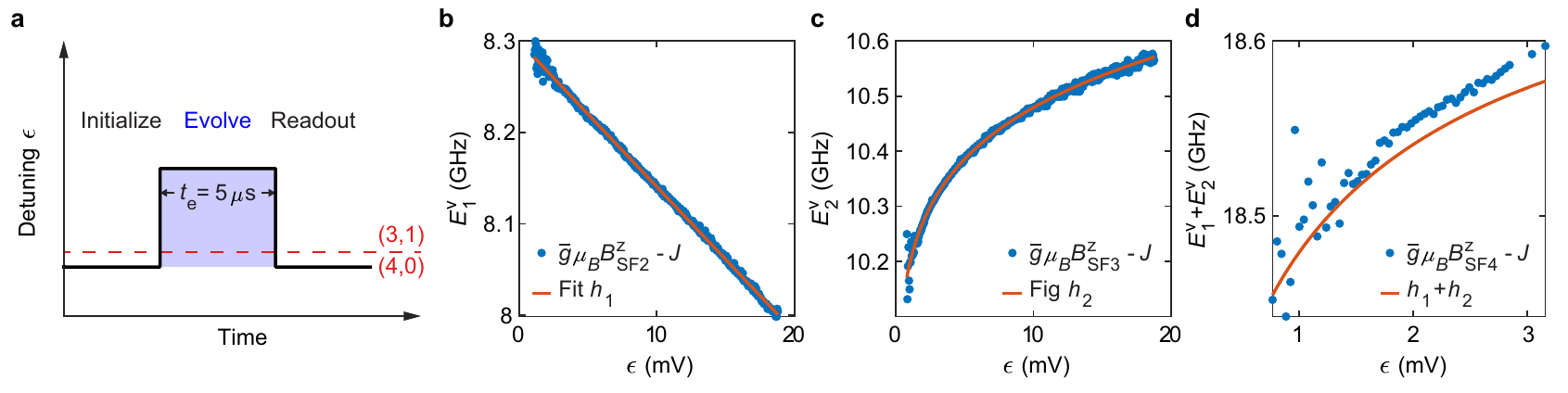}
	\caption{Spin funnel pulse sequence and data analysis.
		\textbf{a}, Pulse sequence used to measure the spin funnels. The system is initialized in the (4,0) singlet ground state, pulsed to variable $\epsilon$ for a fixed time $t_e$, and then pulsed to the PSB region for measurement. The dashed line indicates the (4,0)-(3,1) transition point, where $\epsilon=0$.
		\textbf{b}, Voltage dependence of the valley splitting in dot~1, $\Ev_1$, versus $\epsilon$ extracted from the second spin funnel (blue dots) and fit (red line). 
		\textbf{c}, Voltage dependence of the valley splitting in dot~2, $\Ev_2$, versus $\epsilon$ extracted from the third spin funnel (blue dots) and fit (red line). 
		\textbf{d}, $\Ev_1+\Ev_2$ versus $\epsilon$, extracted from the fourth spin funnel (blue dots) and prediction (red line). The prediction is the sum of the two fits from panels \textbf{b} and \textbf{c}. 
	}
	\label{fig:fitSF}
\end{figure}

We measure spin funnels using the below-described procedure. For variable $B^z$, we prepare the double dot in the (4,0) singlet state and then rapidly pulse $\eps$ into (3,1) for a fixed amount of time $t_e=5~\mu$s (Fig.~\ref{fig:fitSF}a). After this time, we measure the singlet return probability by rapidly pulsing to the Pauli-spin blockade (PSB) region in (4,0). We trace out the spin funnels by extracting the values of $\epsilon$ and $B^z$ at which there are local minima in the singlet return probability. 
For each value of $\eps$, we thus obtain the values of $B^z$ such that $\bar{g} \mu_B B^z = \Delta + J(\epsilon)$, where $\Delta$ can be any of $\{0,E^v_1, E^v_2, E^v_1+E^v_2\}$, depending on the spin funnel at hand. 
We thus obtain four sets of $B^z$ values corresponding to the four different spin funnels, and denote them as $B^z_{^{SFi}}$, where the subscript $_{^{SFi}} (i=1,2,3,4)$ refers to the corresponding spin funnel number.

We determine the voltage dependence of the exchange interaction, $J(\epsilon)$, through two standard techniques. First, we extract $J(\epsilon)$ from the data of the first spin funnel based on the relation $J(\epsilon) = \bar{g}\mu_B B^z_{^{SF1}}$. In the calculations of the Zeeman energy, we assume $\bar{g}=\frac{1}{2}\left(g_1+g_2\right)=2$, the average $g$-factor of dots 1 and 2, is voltage-independent. We find that this method for calibrating $J(\epsilon)$ is accurate for $J(\epsilon) > 10$ MHz. For smaller values of $J(\epsilon)$, the accuracy of this method is limited by the resolution of our magnet.  

\begin{figure}[ht!]
	\includegraphics[width =\textwidth]{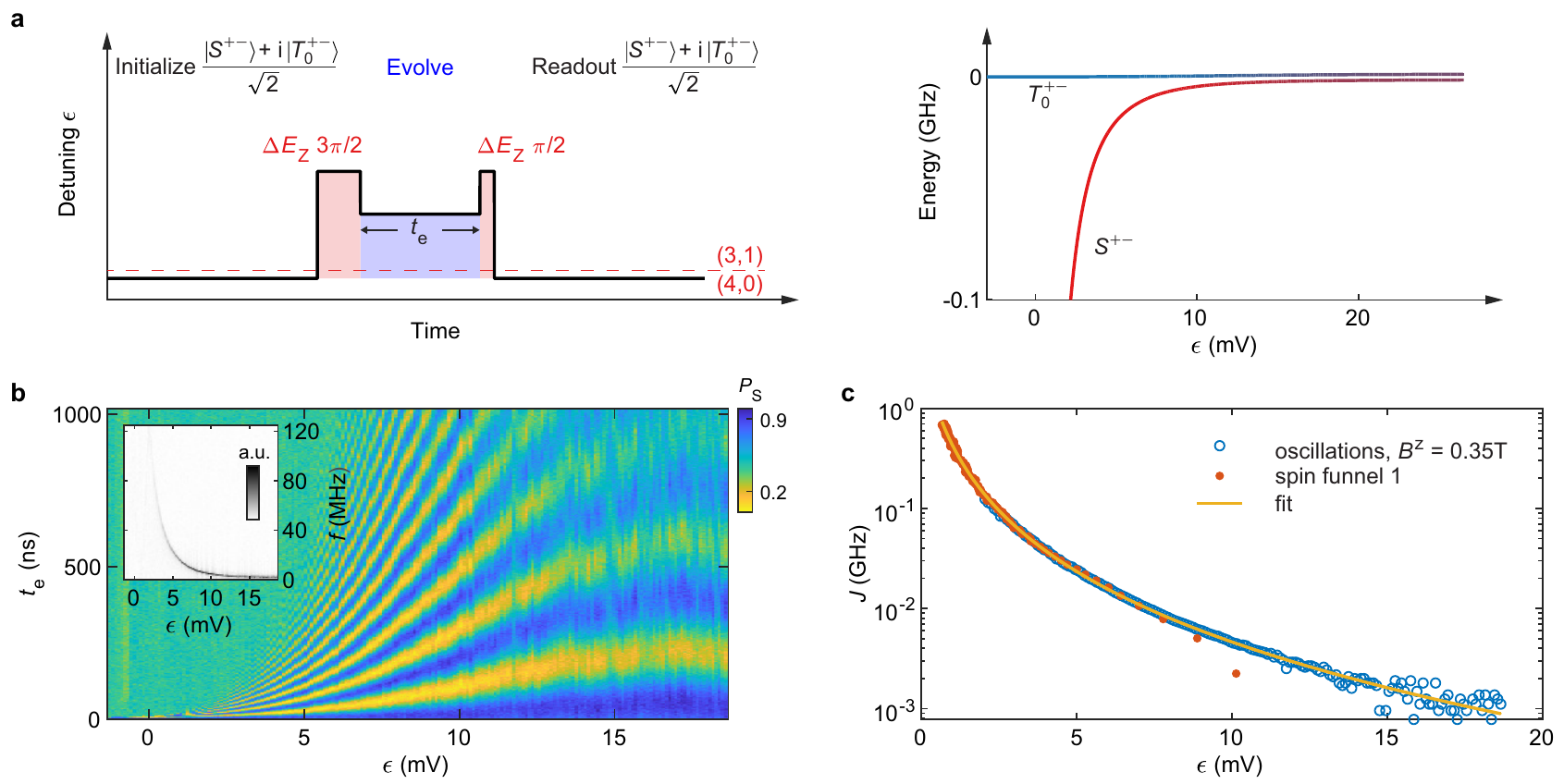}
	\caption{Voltage dependence of the exchange interaction, $J(\epsilon)$.
		\textbf{a}, Pulse sequence used to measure the exchange oscillations in \textbf{b} (left panel) and  schematic of the relevant energy levels (right panel). After initializing the system in the (4,0) singlet ground state, we prepare the superposition state $(\ket{S^{\plus\minus}}+i\ket{T_0^{\plus\minus}})/\sqrt{2}$ via a $3\pi/2$ pulse around $\dez$ at $\epsilon=19$~mV. The state evolves at different values of $\epsilon$ (therefore under different values of $J$) for a variable time $t_e$. After the evolution, we use a $\pi/2$ pulse around $\dez$ to map this superposition state to $\ket{S^{\plus\minus}}$ and measure the singlet return probability $P_S$. 
		\textbf{b}, Measurement of exchange oscillations between superpositions of $\ket{S^{\plus\minus}}$ and $\ket{T_0^{\plus\minus}}$ at different values of $\epsilon$ with $B^z = 350$~mT. Inset: Absolute value of the fast Fourier transform of the data.
		\textbf{c}, Values of $J$ as a function of $\epsilon$ extracted from the exchange oscillation measurement in \textbf{b} (blue circles) and the spin funnel measurement (red dots). The $J$ values are displayed on a logarithmic scale. The solid line is an empirical fit of the data.
	}
	\label{fig:sup_fitJ}
\end{figure}

For smaller values of $J(\epsilon)$, we measure exchange oscillations between superpositions of $\ket{T_0^{\plus\minus}}$ and $\ket{S^{\plus\minus}}$~\cite{Petta2005,Foletti2009,Dial2013,Connors2021charge} (Fig.~\ref{fig:sup_fitJ}). We extract $J(\epsilon)$ from the oscillation frequencies  $f(\epsilon) = \sqrt{J(\epsilon)^2 + (\dez)^2}$, which correspond to the total energy splitting between the eigenstates of the \stz\ qubit Hamiltonian $H_{ST}=(J(\epsilon) \sigma^z+\dez \sigma^x)/2$ at different values of $\epsilon$ (we determine $f(\epsilon)$ from the FFT of the oscillations). To avoid complications arising from \stm\ mixing, the oscillations are measured at $B^z = 350$~mT, away from any of the spin funnels. At this magnetic field, $\dez = 2.52 $~MHz is used to compute $J(\epsilon)$. 

Fig.~\ref{fig:sup_fitJ}c shows the results of the two measurements discussed above, which agree well where they overlap. We find that the data can be well fit by a function of the form $J(\epsilon) = \exp[-p_1 (\epsilon + p_2)^{p_3} + p_4]$, where $p_i$ are the fit parameters. For negative values of $\epsilon$, the energy splitting between $\ket{T_0^{\plus\minus}}$ and the hybridized (4,0) singlet is approximately given by the form $J(-\epsilon) - \alpha\epsilon$, where $\alpha$ is the detuning lever-arm. For the purpose of minimizing the number of fit parameters, we set these two forms to have the same slope at $\epsilon=0$.
By requiring $\partial{J(\epsilon)}/\partial{\epsilon}|_{\epsilon=0} = \partial{(J(-\epsilon)-\alpha\epsilon)}/\partial{\epsilon}|_{\epsilon=0}$, we replace the parameter $p_4$ with $\ln[\alpha/(2p_1 p_3 p_2^{p_3-1})] + p_1 p_2^{p_3}$ in the above function with $\alpha=0.21$~eV/V. We fit the function to the values of $J(\epsilon)$ extracted from the exchange oscillation measurement for $\epsilon>4.5$~mV and the spin funnel measurement for $\epsilon<4.5$~mV simultaneously. The fit result is shown as the solid line in Fig.~\ref{fig:sup_fitJ}c. Setting $\epsilon = 0$ in the function yields an estimate of the interdot tunnel coupling $t_c\sim20~\mu$eV. 
We use the extracted form of $J(\epsilon)$ together with the $\epsilon$ and $B^z$ values corresponding to the second and third spin funnels to calculate the valley splittings in dots 1 and 2 as functions of $\epsilon$.
Specifically, the valley splitting of dot 1 is determined by $E_1^v(\epsilon)=\bar{g}\mu_B B^z_{^{SF2}}-J(\epsilon)$, and the valley splitting of dot 2 is determined by $E_2^v(\epsilon)=\bar{g}\mu_B B^z_{^{SF3}}-J(\epsilon)$.
The extracted values of $\Ev_1$ and $\Ev_2$ are shown in Figs.~\ref{fig:fitSF}b-c. We find that both sets of data can be well fit by a function $h_{1(2)}(\epsilon) = p_1 + p_2\epsilon^{p_3}$, where $p_i (i=1,2,3)$ are fit parameters. 

By fitting to the data in Figs.~\ref{fig:fitSF}b-c and assuming that $E^v_{1(2)}$ is only dependent on the gate voltage $V_{1(2)}$, we find $E^v_1(\Delta V_1)=8.30 - 0.016(-\Delta V_1)^{1.01}$ and $E^v_2(\Delta V_2) =9.16+1.03(\Delta V_2)^{0.11}$ for $\eps$ between 1-19~mV, where $E^v_{1(2)}$ is in units of GHz and $\Delta V_{1(2)}$ in mV. We obtain a valley splitting lever arm of $dE^v_1/d\Delta V_{1}=0.066$~meV/V for dot~1 and values of $dE^v_2/d\Delta V_{2}$ between $0.033-0.54$~meV/V for dot~2. The extracted lever arms are both positive as expected, and the magnitudes are comparable to those of recent reports~\cite{Yang2013,Liu2021Gradient,Jock2021}.
We also show the values of $\bar{g} \mu_B B^z_{^{SF4}}-J(\epsilon)$, obtained for the fourth spin funnel (Fig.~\ref{fig:fitSF}d). We expect that the fourth spin funnel will occur when $\bar{g} \mu_B B^z_{^{SF4}} = E^v_1(\epsilon) +E^v_2(\epsilon)+ J(\epsilon)$.  These values are approximately the sum of our determined values of $\Ev_1$ and $\Ev_2$ with discrepancies of less than 1~mT, indicating that our procedure for extracting the valley splittings is accurate.

\subsection{Magnitude and sign of $\Delta E_Z$}

\begin{figure}[h]
	\includegraphics[width =\textwidth]{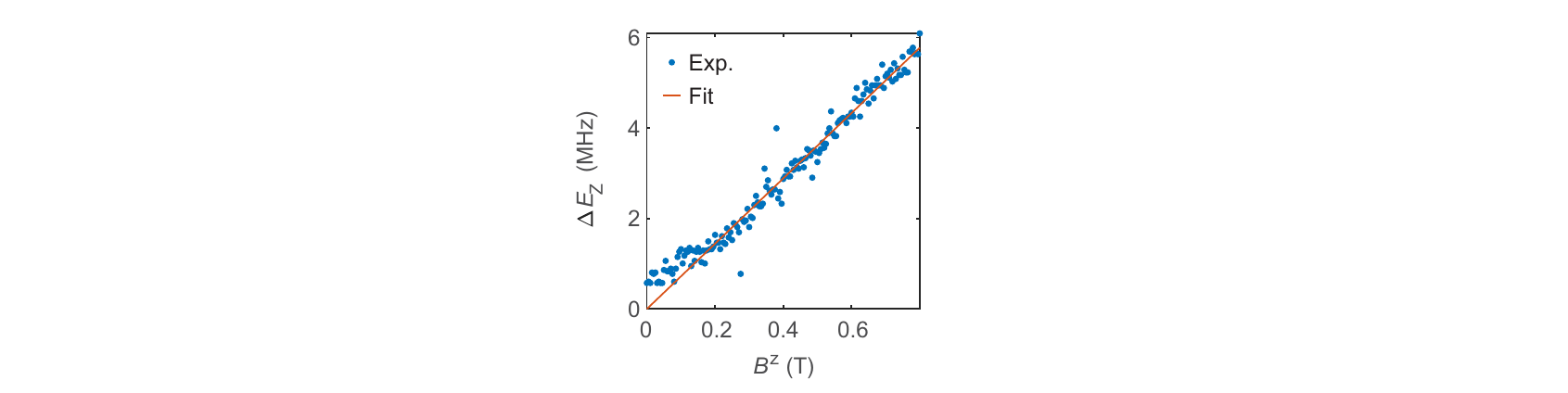}
	\caption{Measured $\dez$ oscillation frequency as a function of external magnetic field $B^z$. Linear fit to the data above 100~mT shows dependence of 7.21~MHz/T, corresponding to a $g$-factor gradient $\Delta g = g_1-g_2 = 5.15\times10^{-4}$. 
	}
	\label{fig:dEzvsBz}
\end{figure}

\begin{figure}[htpb]
	\includegraphics{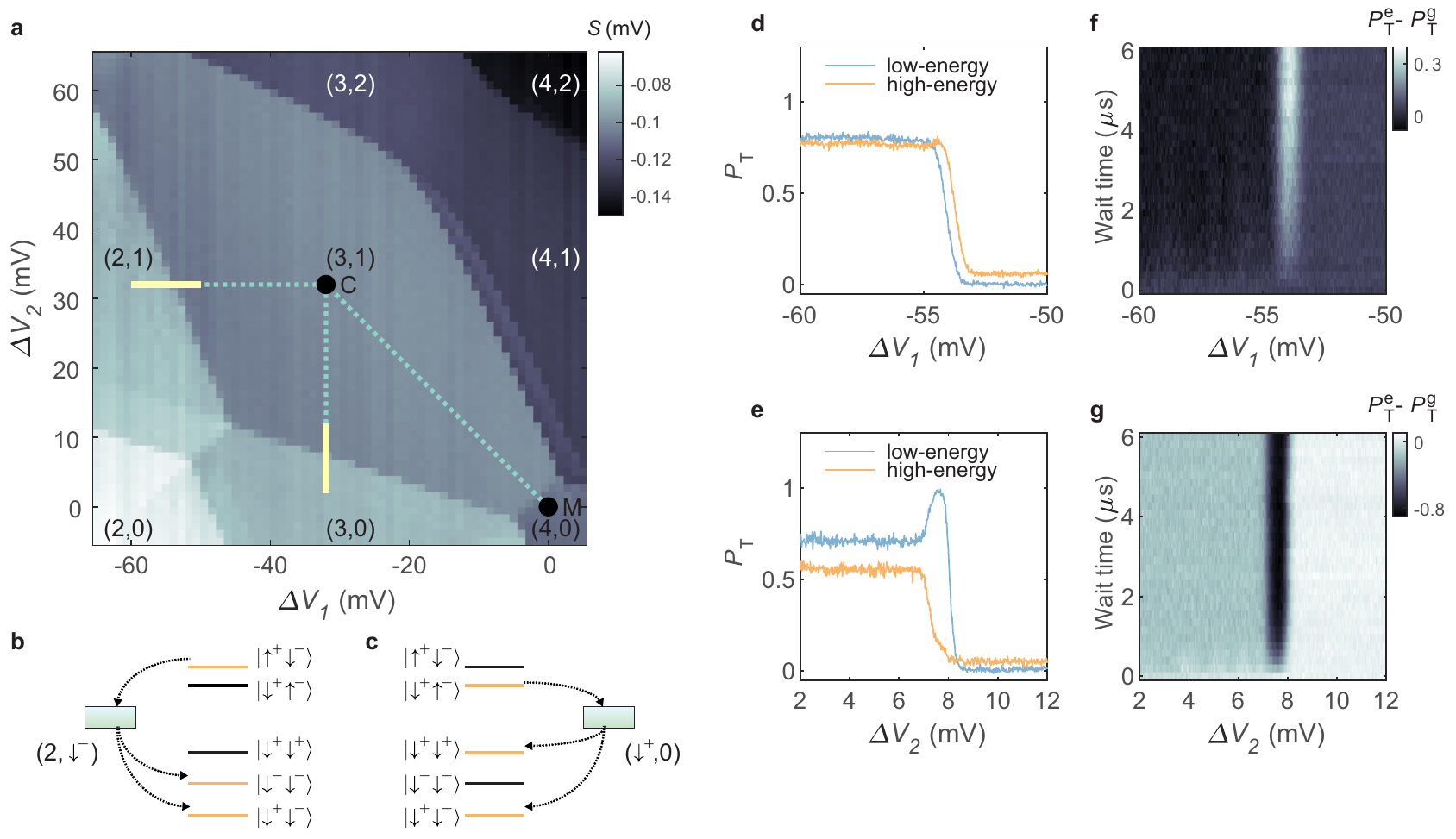}
	\caption{Measurement of the sign of $\dez=g_1\mu_B B^z-g_2 \mu_B B^z$.
		\textbf{a}, Charge stability diagram of the (3,1) charge region illustrating the pulse positions for the relevant measurements. After adiabatically preparing either $\DU$ or $\UD$, we ramp $V_1$ and $V_2$ near either the dot-1 or dot-2 charge transitions (yellow lines) and wait for a variable time. 	
		We then reverse the pulse sequence and use PSB to distinguish between singlet and triplet states.
		\textbf{b}, Schematic of the relevant energy levels during relaxation near the dot-1 transition. 
		Here, $\UD$ may relax to $\ket{\D^{\minus}\D^{\minus}}$ or $\ket{\D^{\plus}\D^{\minus}}$ via exchange with (2,1) states.
		$\DU$ cannot transition to any of $\ket{T_-^{\plus\minus}}$, $\ket{T_-^{\minus\minus}}$, or $\ket{T_-^{\plus\plus}}$ without undergoing, at minimum, a spin flip in dot 2 and therefore does not appreciably relax during the microsecond timescales in these measurements.
		\textbf{c}, Schematic of the relevant energy levels during relaxation near the dot-2 transition. Here, $\DU$ may relax to $\ket{\D^{\plus}\D^{\plus}}$ or $\ket{\D^{\plus}\D^{\minus}}$ via exchange with (3,0) states.
		$\UD$ cannot transition to $\ket{T_-^{\plus\minus}}$, $\ket{T_-^{\minus\minus}}$, or $\ket{T_-^{\plus\plus}}$ without incurring a spin flip in dot 1, and therefore does not appreciably relax. 
		\textbf{d}, Triplet return probability near the dot-1 transition, corresponding to \textbf{b}. 
		When preparing the higher-energy state, there is a slight enhancement in the triplet return probability inside the (3,1) region, suggesting that we have prepared the state $\UD$.
		\textbf{e}, Triplet return probability near the dot-2 transition, corresponding to \textbf{c}. 
		When preparing the lower-energy state, there is a strong enhancement in the triplet return probability inside the (3,1) region, indicating that we have prepared the state $\DU$.
		The overall visibility of the higher-energy state traces is lower due to imperfections in the preparation and readout.
		We suspect that the stronger enhancement in the triplet return probability observed near the dot 2 transition compared to the enhancement observed near the dot 1 transition in \textbf{d} may be due to the details affecting the relaxation rates between the  states involved in the relaxation processes.
		\textbf{f}, Plot of the triplet return probability for the high-energy-state measurement, $P^e_T$, minus the triplet return probability of the low-energy-state measurement, $P^g_T$, as a function of the wait time and wait position near the dot-1 transition.
		\textbf{g}, Plot of $P^e_T - P^g_T$ as a function of the wait time and wait position  near the dot-2 transition.
		From the data shown in \textbf{d}-\textbf{g}, we conclude that the low-energy state is $\DU$, and therefore $\dez>0$.
		All data shown in this figure are acquired at $B^z=600~\text{mT}$.
		}\label{fig:sup_dbzDir}
\end{figure}

In this work, we define the difference in the Zeeman energy between the two dots as $\dez\equiv E_1^z-E_2^z=g_1\mu_B B^z-g_2\mu_B B^z$. We have neglected the hyperfine fields, because we find that the $g$-factor difference is the dominant mechanism for $\dez$~\cite{kerckhoff2021magnetic,Connors2021charge}.
We measure the magnitude of $\dez$ through coherent oscillations at large $\epsilon$ using a pulse sequence similar to Fig.~\ref{fig:sup_STMosc}a. With the state $\ket{S^{\plus\minus}}$ initialized, we rapidly pulse to $\epsilon=34$~mV, wait for a variable evolution time, and then rapidly pulse back to (4,0) to measure $P_S$. In agreement with previous work \cite{Connors2021charge,Eng2014,kerckhoff2021magnetic}, we observe that the $\dez$ oscillation frequency increases linearly with the external field $B^z$ for $B^z>200$~mT (Fig.~\ref{fig:dEzvsBz}). A linear fit to the data yields a slope   $\dez/B^z=7.21$~MHz/T, corresponding to a $g$-factor gradient between the two dots of $\Delta g = g_1-g_2 = 5.15\times10^{-4}$.

We next determine the sign of $\dez$. Whether this quantity is positive or negative determines whether or not $\ket{S^{\plus\minus}}$ maps to $\UD$ or $\DU$ after adiabatic separation of the valence electrons.  
In order to determine the sign of $\Delta E_Z$, we measure the relaxation of both $\UD$ and $\DU$ to different $\tm$ states near the (3,1)-(2,1) and (3,1)-(3,0) charge transitions via the procedure described in Ref.~\cite{Orona2018} and below. 
For all of the measurements described in this section, $B^z=600$~mT, such that $\ket{T_-^{\plus\minus}}$, $\ket{T_-^{\minus\minus}}$, and $\ket{T_-^{\plus\plus}}$ are lower in energy than $\UD$ and $\DU$.

All measurements begin by initializing a (4,0) singlet state via electron exchange between dot 1 and its corresponding reservoir in the (4,0) charge region. 
To prepare the lower-energy spin-zero eigenstate of $\dez$ (either $\DU$ or $\UD$), we adiabatically separate the electrons by ramping $\eps$ adiabatically from position M to position C in Figure~\ref{fig:sup_dbzDir}a~\cite{Petta2005,Foletti2009}. 
We then ramp the double-dot chemical potentials near either the (3,1)-(2,1) or (3,1)-(3,0) charge transitions, such that either dot 1 or dot 2 can exchange an electron with its reservoir, but not both. Then, we wait for a variable amount of time, and then ramp the chemical potentials back to position C.
Finally, we adiabatically ramp $\epsilon$ back to position M for a PSB readout to determine whether the final state is a singlet or a triplet.
To conduct a similar measurement with the higher-energy spin-zero eigenstate of $\dez$ state, we apply the exact same pulse sequence, but we apply $\dez$ $\pi$ pulses immediately after (4,0) singlet initialization and immediately after ramping back to position M. 

If the electron in dot 1(2) has spin up after the initial ramp to point C, then during the wait near the dot 1(2) charge transition, the electron in dot 1(2) may tunnel out and a spin-down electron may tunnel into that dot, yielding an increased triplet return probability (Figs.~\ref{fig:sup_dbzDir}b-c). 
Thus, we expect to see an increase in the triplet return probability near one of the charge transitions when we prepare the lower-energy spin-zero eigenstate of $\dez$, and near the opposite charge transition when we prepare the higher-energy spin-zero eigenstate of $\dez$. 

Specifically, in the case where we have prepared $\UD\left(\DU\right)$, we expect to see an enhancement in the triplet return probability near the dot 1(2) transition line. 
Figures~\ref{fig:sup_dbzDir}d-g show plots of the triplet return probability near the two relevant charge transitions for both the lower- and higher-energy states.
When the higher-energy spin-zero eigenstate of $\dez$ is prepared, a clear region of enhanced triplet return probability is visible when the wait position is near the dot 1 transition (Fig.~\ref{fig:sup_dbzDir}d), suggesting that lower-(higher-) energy state is $\DU$($\UD$).

We corroborate this by observing a similar enhancement of the triplet return probability when we prepare the lower-energy state, and when the wait position is near the dot 2 transition (Fig.~\ref{fig:sup_dbzDir}e). 
Thus, we conclude that upon adiabatic separation of the valence electrons, $\ket{S^{\plus\minus}}$($\ket{T_0^{\plus\minus}}$) transitions to $\DU$($\UD$), and that $\dez$ is positive. 

These data, together with the data of Fig.~3 in the main text, enable us to assign valley splittings to the different dots. Because we observe single-spin Rabi oscillations on the second spin funnel when we prepare the higher-energy spin-zero eigehstate of $\dez$ ($\UD$), we assign the valley splitting associated with the second spin funnel to dot 1. Likewise, we assign the valley splitting associated with the third spin funnel to dot 2.

\subsection{Single-spin-valley oscillations}

\begin{figure}[t!]
	\includegraphics[width =\linewidth]{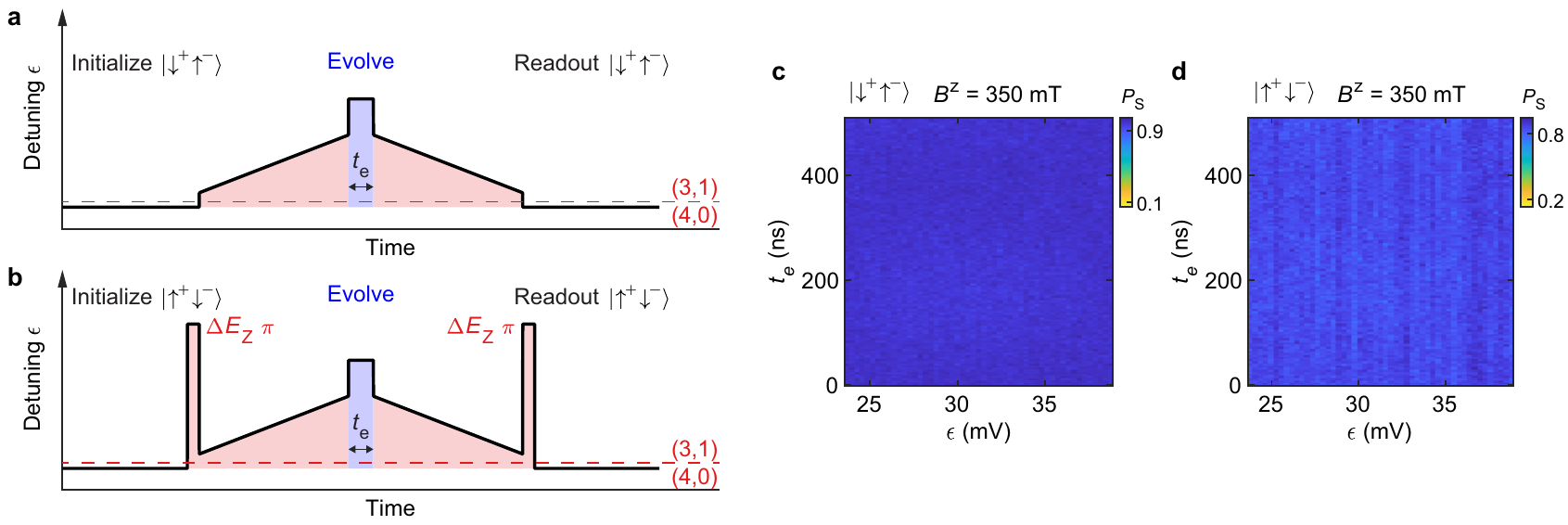}
	\caption{Single-spin Rabi oscillation pulse sequence and control measurements.
		\textbf{a} We prepare the state $\DU$ by adiabatically separating a singlet state in the presence of $\dez$. 
		After an evolution period of variable time, $t_e$, we reverse the initialization process to map $\DU$ back to $\ket{S^{\minus\plus}}$ for PSB readout. 
		\textbf{b} We prepare $\UD$ in a similar way, except that we add a $\dez$ $\pi$ pulse before separating and after recombining the electrons. The initial $\pi$ pulse rotates the state $\ket{S^{\plus\minus}}$ to $\ket{T_0^{\plus\minus}}$. The $\ket{T_0^{\plus\minus}}$ state, in turn, evolves to the state $\UD$ upon adiabatic separation. 
		The readout follows the same steps in reverse, where the adiabatic pulse maps $\UD$ back to $\ket{T_0^{\plus\minus}}$ and then the $\pi$ pulse maps $\ket{T_0^{\plus\minus}}$ to $\ket{S^{\plus\minus}}$. 
		\textbf{c,d} Control measurements at a magnetic field away from the spin funnels, $B^z=350$~mT, using the pulse sequence in \textbf{a} and the pulse sequence in \textbf{b}, respectively. No oscillations are observed in either case.
	}
	\label{fig:sup_hotspot}
\end{figure}

\begin{figure}[t!]
 	\includegraphics[width =\linewidth]{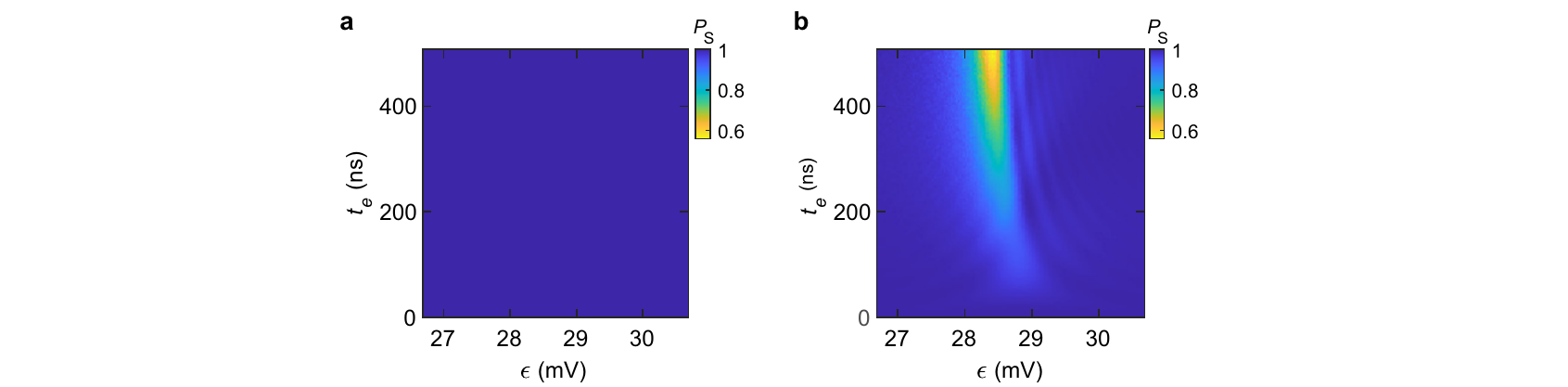}
	\caption{Numerical simulations corresponding to Fig.~3c in the main text. In both cases, the state is prepared and measured along $\tildeDU$.
		\textbf{a} Simulation assuming $J_r=0$ and $\eta=0$.
		\textbf{b} Simulation assuming $J_r=0.5$~MHz and $\eta=0.0167$.
	}
	\label{fig:sup_hotspot_sim}
\end{figure}

Figures~\ref{fig:sup_hotspot}a-b show the pulse sequences used to acquire the data shown in Fig. 3 in the main text corresponding to single-spin-valley oscillations (or lack thereof).
We begin by initializing the system in the (4,0) singlet state, from which we prepare $\DU$ via adiabatic separation of the electrons by ramping $\epsilon$ to deep into (3,1), and we then pulse to different values of $\epsilon$ near the center of (3,1).
After a variable evolution time, we measure the $\DU$ component of the state 
via a reverse process to map $\DU$ back to $\ket{S^{\plus\minus}}$ for PSB readout.
To conduct a similar experiment with the $\UD$ state, we apply a $\pi$ pulse about $\dez$  immediately before electron separation and immediately after recombination. 

Figures~\ref{fig:sup_hotspot}c-d show measurements similar to those in Fig.~3 in the main text, except that the magnetic field is tuned away from any of the spin funnels. As expected, there are no distinguishable features in these data, confirming that the oscillations observed in Fig.~3 in the main text are related to the spin-valley coupling.

To understand the features in Figs.~3c and 3e in the main text, we investigate the impact of the residual exchange, $J_r$, via numerical simulations. 
Its effects are two-fold. 
First, it couples $\UD$ to $\DU$. 
Second, with non-zero $J_r$, the spin-zero eigenstates deep in (3,1) are not exactly the product states $\DU$ and $\UD$, but are instead given by
\begin{align}
    \tildeDU=& \sqrt{1-\eta}\DU - \sqrt{\eta}\UD
\end{align}
and 
\begin{align}
    \tildeUD=& \sqrt{1-\eta}\UD + \sqrt{\eta}\DU,
\end{align}

\noindent where $\eta$ is a small, positive number, the magnitude of which is determined by $J_r$ and $\dez$.
This results in the adiabatic state preparation (and readout) processes described above (Figs.~\ref{fig:sup_hotspot}a-b) mapping $\ket{S^{\plus\minus}}$ and $\ket{T_0^{\plus\minus}}$, not perfectly to $\DU$ and $\UD$ as intended, but to $\tildeDU$ and $\tildeUD$, respectively.
Fig.~\ref{fig:sup_hotspot_sim} shows the numerical simulations corresponding to Fig.~3c of the main text, where we consider $J_r=0$ and $J_r=0.5~$MHz.
The simulation that incorporates non-zero $J_r$ (Fig.~\ref{fig:sup_hotspot_sim}b) captures the main features in Fig.~3c. 
Spurious coupling between the states $\DU$ and $\ket{\D^{\minus} \D^{\minus}}$ or between $\UD$ and $\ket{\D^{\plus} \D^{\plus}}$ may also result in similar features to those that we observe in Fig.~3c or 3e. However, such transitions, which would require a pure spin flip in one dot and a pure valley change in the other, are expected to have a much lower probability to occur because both the transverse hyperfine gradient and the intervalley coupling are expected to be small. 
The presence of a weak transverse gradient explains why the first spin funnel is relatively dim, as discussed in the main text, while weak intervalley coupling is responsible for the dimness of the fourth spin funnel. 

\subsection{\stm\ Rabi, Ramsey, and spin-echo measurements}


\begin{figure}[h!]
	\includegraphics[width =\linewidth]{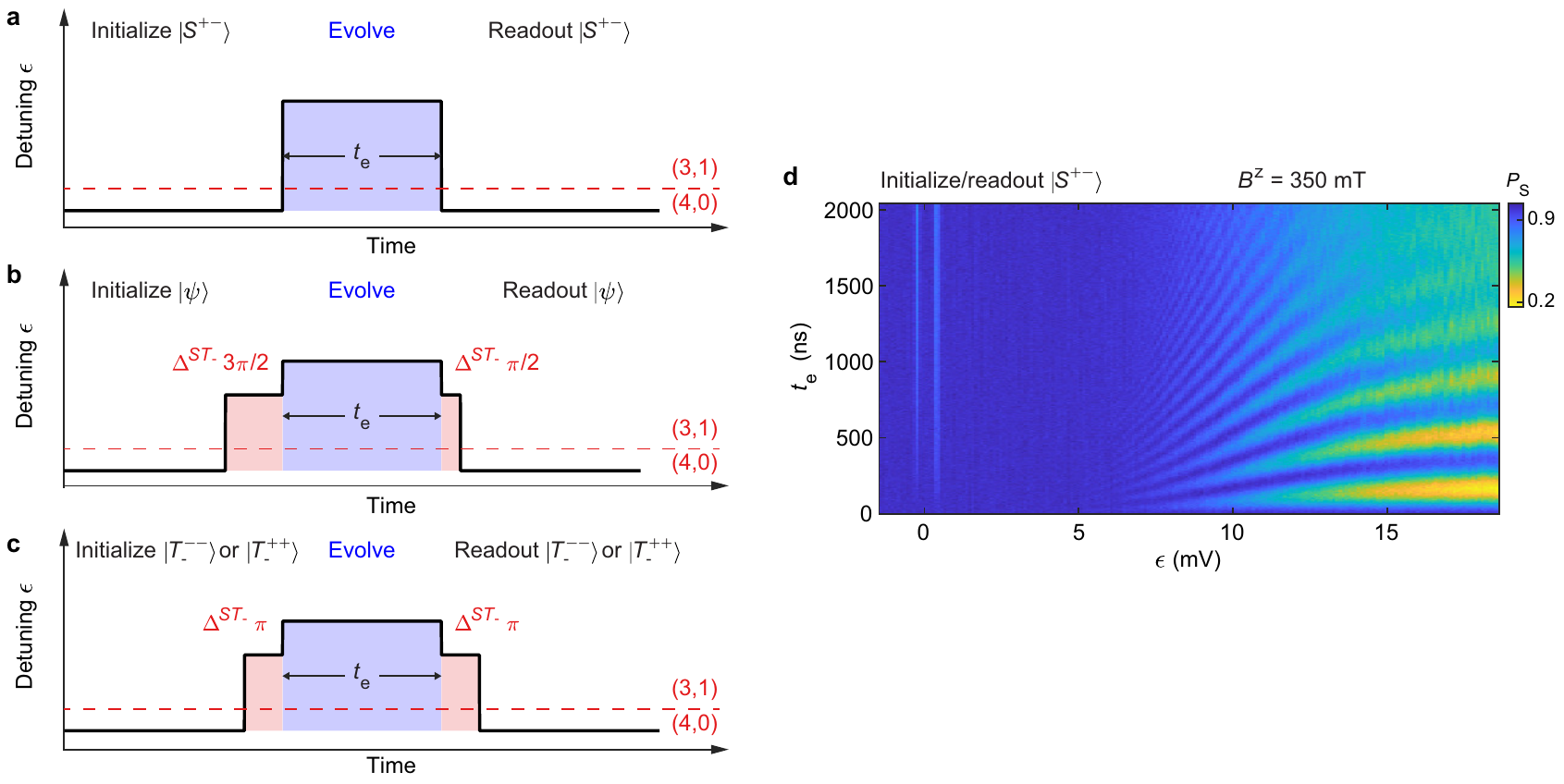}
	\caption{Pulse sequences for coherent spin-valley-driven singlet-triplet oscillations. 
		\textbf{a} Pulse sequence for \stm\ Rabi oscillations. We prepare the double dot in the state $\ket{S^{\plus\minus}}$, pulse to different values of $\epsilon$ for a variable evolution time $t_e$, and then measure the singlet return probability $P_S$ by pulsing $\epsilon$ to the PSB region in (4,0).
		\textbf{b} Pulse sequence for \stm\ Ramsey oscillations, with a $3\pi/2$ and $\pi/2$ pulse performed at the $\epsilon$ value of the \stm{} avoided crossing (denoted $\Delta^{S\tm}$) before and after the evolution. Through a $\Delta^{S\tm}$ $3\pi/2$ pulse, we prepare the double dot in a superposition of $\ket{S^{\plus\minus}}$ and either $\ket{T_-^{\minus\minus}}$ or $\ket{T_-^{\plus\plus}}$, depending on which spin funnel we are operating near. 
		Specifically, for the second spin funnel, we prepare $\ket{\psi} = \frac{1}{\sqrt{2}}(\ket{S^{\plus\minus}} + i\ket{\tm^{\minus\minus}})$ and for the third spin funnel, we prepare $\ket{\psi} = \frac{1}{\sqrt{2}}(\ket{S^{\plus\minus}} - i\ket{\tm^{\plus\plus}})$, if $\dsv_{1(2)}$ are real and positive. After the evolution, we use a $\Delta^{S\tm}$ $\pi/2$ pulse to map $\ket{\psi}$ to $\ket{S^{\plus\minus}}$ for PSB readout.
		\textbf{c} Pulse sequence used to observe triplet-triplet oscillations, with a $\pi$ pulse at the $\epsilon$ value of the \stm{} avoided crossing before and after the evolution. Through a $\Delta^{S\tm}$ $\pi$ pulse, we prepare the excited $\ket{\tm}$ state, $\ket{\tm^{\minus\minus}}$ for the second spin funnel or $\ket{\tm^{\plus\plus}}$ for the third spin funnel. After the evolution, we apply another $\Delta^{S\tm}$ $\pi$ pulse to map the excited $\ket{\tm}$ state to $\ket{S^{\plus\minus}}$ for PSB readout. 
	    \textbf{d} Control measurement at a magnetic away from the spin funnels, $B^z=350$~mT, using the pulse sequence in \textbf{a}. 
	    The energy level diagram for this magnetic field is displayed in Fig.~\ref{fig:sup_fitJ}a, right panel.
	    The two vertical lines in the data near $\eps=0$ correspond to where $\ket{T_-^{\plus\minus}}$ and $\ket{T_-^{\minus\minus}}$ come into resonance with $\ket{S^{\plus\minus}}$, which occur at energies below the range plotted in Fig.~\ref{fig:sup_fitJ}a.
	}
	\label{fig:sup_STMosc}
\end{figure}	

\begin{figure}[h!]
	\includegraphics[width =\linewidth]{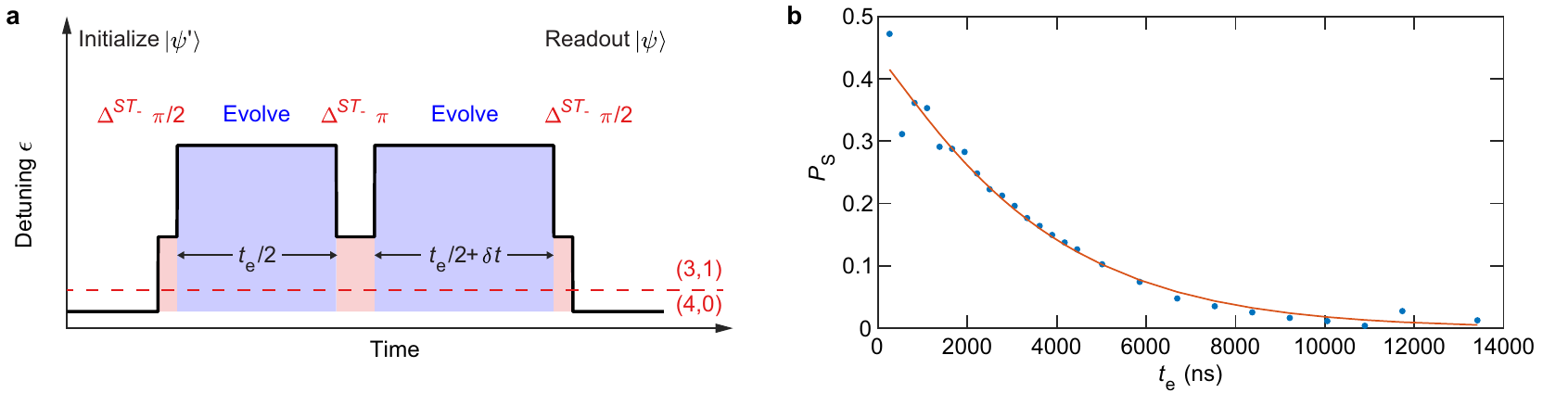}
	\caption{\stm\ spin echo pulse sequence and data.
		\textbf{a} The double dot is prepared in a superposition of $\ket{S^{\plus\minus}}$ and $\ket{\tm^{\plus\plus}}$ via a $\pi/2$ pulse at the \stm\ avoided crossing on the third spin funnel. Specifically, we prepare $\ket{\psi'}= \frac{1}{\sqrt2}(\ket{S^{\plus\minus}} + i \ket{\tm^{\plus\plus}})$ for real, positive $\dsv_2$. The state evolves near  $\epsilon=9$~mV, where the \stm splitting is approximately 50~MHz and varies roughly linearly with $\eps$, for a total time $t_e +\delta t$, during which we apply a $\pi$ pulse at the avoided crossing to refocus the dephasing. After the evolution is complete, we apply another $\pi/2$ pulse before PSB readout. 	
		\textbf{b} Echo amplitude decay as a function of the total qubit evolution time $t_e$. To determine the amplitude at each point, we fit the inhomogeneously broadened decay for fixed $t_e$ and varying $\delta t$~\cite{Dial2013,Connors2021charge}. We find $T_2^*\approx 210$~ns and $T_2^{echo} = 3.6~\mu$s. 
	}
	\label{fig:sup_echo}
\end{figure}

The Rabi oscillation measurements in Fig.~4 in the main text were obtained using the pulse sequence of Fig.~\ref{fig:sup_STMosc}a. A control measurement corresponding to Fig.~4a in the main text appears in Fig.~\ref{fig:sup_STMosc}d. For the control measurement, the magnetic field was tuned away from any of the spin funnels ($B^z=350$~mT), and no \stm\ oscillations are visible.

The Ramsey oscillation measurements of Fig.~5a in the main text were obtained using the pulse sequence of Fig.~\ref{fig:sup_STMosc}b. The measurements of Fig.~5b in the main text, which highlight the triplet-triplet oscillations, were obtained with the pulse sequence of Fig.~\ref{fig:sup_STMosc}c. 

The Ramsey experiments of Fig.~6 in the main text were also acquired using the pulse sequence shown in Fig.~\ref{fig:sup_STMosc}b. 
For the oscillations measured on the ``left" (smaller $\eps$) \stm{} avoided crossing (Fig.6a), we perform the $3\pi/2$ and $\pi/2$ pulses on the left avoided crossing, whereas for the oscillations measured at the ``right" (larger $\eps$) \stm{} avoided crossing (Fig.6b) and the experiment deep in (3,1) (Fig.6c), the $3\pi/2$ and $\pi/2$ pulses are performed on the right avoided crossing.

As a further demonstration of full (two-axis) control of the \stm\ qubit, we perform a spin-echo dynamical-decoupling protocol in which we extend the coherence time of the spin-valley-driven Ramsey oscillations via a refocusing $\pi$ pulse (Fig.~\ref{fig:sup_echo}a). 
In doing so, we extend the coherence time by a little over an order of magnitude, which is in line with recently reported spin-echo protocols implemented in Si-based qubits to filter out low-frequency charge noise~\cite{Connors2021charge,Jock2018}.
Fig.~\ref{fig:sup_echo}b shows a plot of the echo amplitude decay as a function of the total qubit evolution time, measured for the case of a single set of \stm\ Rabi oscillations on the third spin funnel.

\subsection{$\dsv$ dependence on the orientation of the magnetic field}

\begin{figure}[htbp]
	\includegraphics[width =\textwidth]{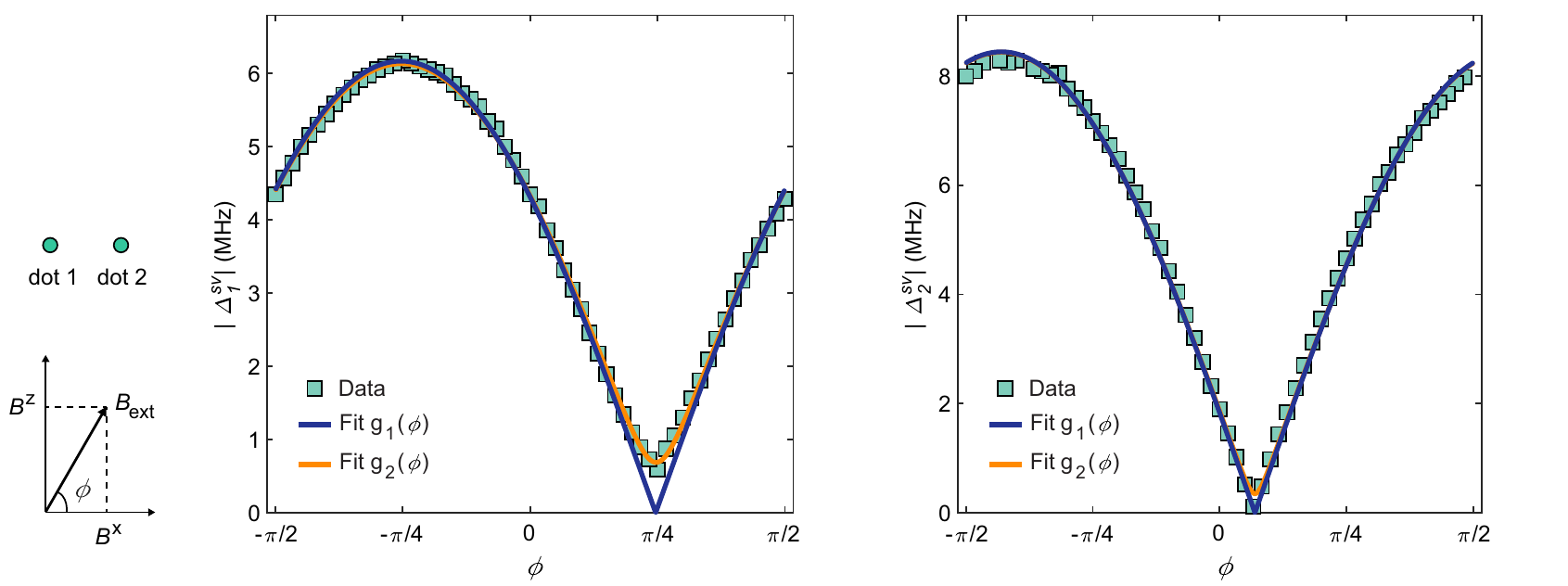}
	\caption{Measured magnitudes of $\dsv_1$ and $\dsv_2$ versus the in-plane magnetic field angle $\phi$ (blue squares). The amplitude of the magnetic field is fixed at $|B_{ext}|=\sqrt{(B^x)^2+(B^z)^2}=300$~mT in \textbf{a} and 375~mT in \textbf{b}, where the $B^x$ component is parallel with the axis connecting the two quantum dots and the $B^z$ component is perpendicular to the axis.
	We fit both sets of data to an equation of the form $g_1(\phi) = |A\sin\phi + B\cos\phi| $, where $A$ and $B$ are fit parameters (dark blue lines), and another equation of the form $g_2(\phi) = |A\sin\phi + B e^{i \theta}\cos\phi|$, where $A$, $B$ and $\theta$ are fit parameters and constrained to be real numbers (orange lines).	
	}\label{fig:sup_angleDep_osc}
\end{figure}

To confirm that the spin-valley coupling arises from spin-orbit coupling, we measured the dependence of the \stm{} oscillation frequency as a function of the in-plane magnetic field angle.  We fix the amplitude of the external magnetic field, $|B_{ext}|= 300$~mT for the second spin funnel and $|B_{ext}|= 375$~mT for the third spin funnel, and vary the orientation of the magnetic field in the plane of the Si/SiGe heterostructure. We use a pulse sequence similar to that shown in Fig.~\ref{fig:sup_STMosc}a, and we record a data set similar to the data of Fig.~4 in the main text for different orientations of the external magnetic field. For each orientation, we locate the $\epsilon$ value of the avoided crossing by finding the largest amplitude of the observed oscillations. Then, we extract the frequency of those oscillations by finding the frequency corresponding to the peak in the fast Fourier transform of the time series at that $\epsilon$ value. We multiply the frequencies by $1/\sqrt2$ to provide an estimate of the spin-valley matrix elements, $\dsv_{1}$ and $\dsv_2$. These data are shown in Fig.~\ref{fig:sup_angleDep_osc}.

We fit our data following the analysis of Ref.~\cite{zhang2020anisotropy}. In brief, Rashba and Dresselhaus spin-orbit coupling can mix single-spin states with different spin and valley quantum numbers. The effective spin-valley matrix elements for a single spin can be represented as $\braket{\U^{\minus}|H^{sv}|\D^{\plus}}=\braket{\U|\bm{B^{so}}\cdot\bm{\sigma}|\D}$ and $\braket{\U^{\plus}|H^{sv}|\D^{\minus}}=\braket{\U|\bm{(B^{so})^*}\cdot\bm{\sigma}|\D}$, where $\ket{\U}$ and $\ket{\D}$ are single-spin eigenstates of the physical magnetic field $\bm{B}\cdot\bm{\sigma}$, $\bm{B^{so}}$ is the effective spin-orbit field, and $\bm{\sigma}$ is a vector of Pauli matrices acting on the spin states. Here, only the first of the two matrix elements is relevant to our study, which we have defined as $\dsv$ (Eq.~\ref{equ:dsv}). According to Ref.~\cite{zhang2020anisotropy}, $\bm{B^{so}}$ lies in the $x$-$y$ plane of the crystal, where the $\hat{x}$ direction corresponds to the $[110]$ direction, and the $\hat{y}$ direction corresponds to the $[\bar{1} 1 0]$ direction. 

Although we do not know the the exact orientation of our device with respect to these crystal directions, the device is fabricated on the $(001)$ plane, and the axis connecting the two quantum dots is approximately aligned with either the $[110]$ or the $[\bar{1}10]$ direction. Thus, the $\hat{x}$ and $\hat{y}$ crystallographic directions lie in the plane of the device, and we expect that $\bm{B^{so}}$ also lies in the plane of our device, and its exact orientation depends on the relative sizes of the Rashba and Dresselhaus contributions.

Assuming the components of $\bm{B^{so}}$ along different crystallographic directions are in phase, such that $B^{so}_{[110]}/B^{so}_{[\bar{1}10]} = Re^{i\theta}$ and $\theta =0$, we expect that the absolute value of the spin-valley matrix element should be described by a function of the form $|\dsv_i(\phi)|\propto|\sin(\phi+\phi_i)|$, where $\phi_i$ depends on the size of the Rashba and Dresselhaus terms in dot $i$. We find that our data are in general well described by this function, as shown by the dark blue lines in Fig.~\ref{fig:sup_angleDep_osc}. 
We notice that the data of $\dsv_1$ seem to suggest a non-vanishing phase difference ($\theta\neq0$) in dot~1, which would mean that $|\dsv_1|$ is non-zero for all values of $\phi$. A non-zero $\theta$ corresponds to a phase difference between the intervalley dipole matrix elements along different crystallographic directions~\cite{zhang2020anisotropy}. 
By including $\theta$ in the fit using the existing model, we find $|\theta|\approx13^\circ$ for dot~1 (Fig.~\ref{fig:sup_angleDep_osc}). 

\subsection{Simulation}
All numerical simulations involve integrating the Schr\"odinger equation with a time-independent Hamiltonian $H^i$, corresponding to spin-valley flips in dot $i$. In the basis $\{\ket{\tm'},\ket{S^{\plus\minus}},\ket{T_0^{\plus\minus}}\}$, where $\ket{\tm'}=\ket{\tm^{\minus\minus}}(\ket{\tm^{\plus\plus}})$ for dot 1(2), 
\begin{equation}
	H^i = \left( \begin{array}{ccc}
		-\bar{E}_Z+E^v_i(\Delta V_i+\delta \Delta V_i)	&	(-1)^{i+1}\dsv_i/\sqrt{2} 	&	\dsv_i/\sqrt{2} \\
		(-1)^{i+1}\dsv_i/\sqrt{2}	&	-J(\eps +\delta \eps)	&	\dez/2 \\
		(\dsv_i)^*/\sqrt{2}	&	\dez/2	&	0
	\end{array} \right).
\end{equation}
Here $\bar{E}_Z=\frac{1}{2}\left(g_1\mu_B B^z_1+ g_2\mu_B B^z_2\right)$ is the average Zeeman energy and $\dez=g_1\mu_B B^z_1-g_2\mu_B B^z_2$ is the Zeeman gradient. $B^z_1=B^z_{ext}+\delta B^{z}_{1}$, and $B^z_2=B^z_{ext}+\delta B^{z}_{2}$, where $\delta B^{z}_{1}$ and $\delta B^{z}_{2}$ are the $z$-components of the hyperfine magnetic fluctuations in dots 1 and 2. We simulate these magnetic fluctuations in our experiments by assuming that $\delta B^z_i$ are independent, quasi-static, Gaussian-distributed random variables with standard deviation $\sigma_B=8~\mu$T, corresponding to an inhomogeneously broadened single-spin coherence time of $1~\mu$s (or a coherence time of $\dez$ rotations of approximately $710$ ns), in good agreement with our observations. For the $g$-factor values, we use $\bar{g}=\frac{1}{2}(g_1+g_2)=2$ and $\Delta g= g_1-g_2 = 5.15\times10^{-4}$ as discussed previously.

The valley splittings $E^v_i (\Delta V_i)$ and exchange coupling $J(\eps)$ are those discussed previously (Figs.~\ref{fig:fitSF}-\ref{fig:sup_fitJ}). We incorporate electrical noise in our simulations by generating independent, quasi-static, Gaussian-distributed voltage noise for the two plunger gates $\delta \Delta V_i$ with empirically-arrived-at rms values of 34 $\mu$V for dot 1 and and 92 $\mu$V for dot 2. To incorporate electrical noise in the detuning, we define fluctuations in the detuning $\delta \epsilon=\delta \Delta V_2 - \delta \Delta V_1$.

We set $\dsv_{1} = 4.1$~MHz and $\dsv_{2} = 7.6$~MHz, as extracted from the single-spin-valley oscillation measurements in the main text. For simplicity, we assumed both $\dsv_{i}$ are real and positive numbers in the simulations. Adding a phase to $\dsv_{i}$ does not affect the simulation results. The pre-factor of $(-1)^{i+1}$ in the \stm{} matrix elements reflects the fact that the sign of the \stm{} spin-valley matrix element changes sign depending on the dot (whereas the \tztm{} spin-valley matrix element does not), as discussed in the main text.

\begin{figure}[htbp]
	\includegraphics[width =\textwidth]{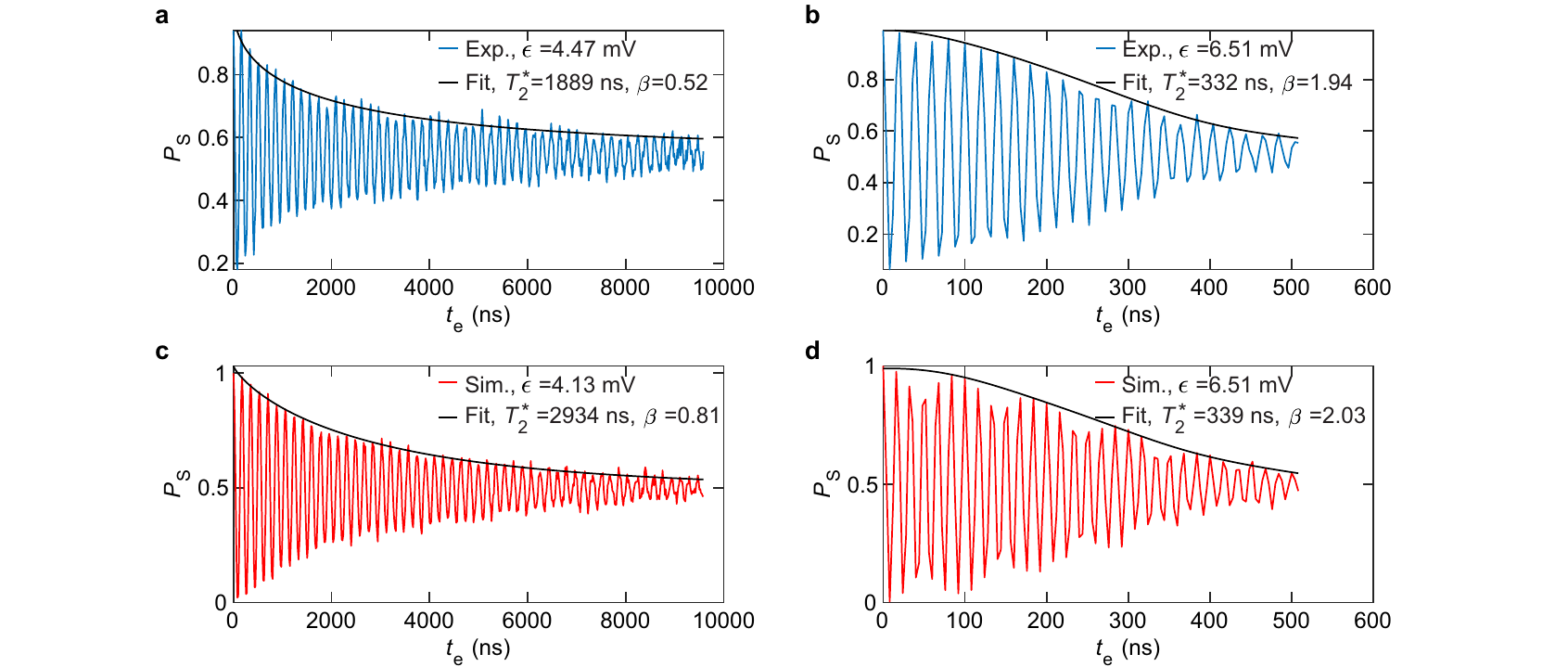}
	\caption{Measured and simulated oscillations near the second spin funnel.
	\textbf{a} Measured Rabi oscillations at the $S^{\plus\minus}$-$T_-^{\minus\minus}$ avoided crossing and envelope of the corresponding fit. 
	\textbf{b} Measured Ramsey oscillations near $\eps=6.5$ mV and envelope of the corresponding fit.
	\textbf{c} Simulation and fit corresponding to panel \textbf{a}.
	\textbf{d} Simulation and fit corresponding to panel \textbf{b}.
	}\label{fig:sup_sim2}
\end{figure}

\begin{figure}[htbp]
	\includegraphics[width =\textwidth]{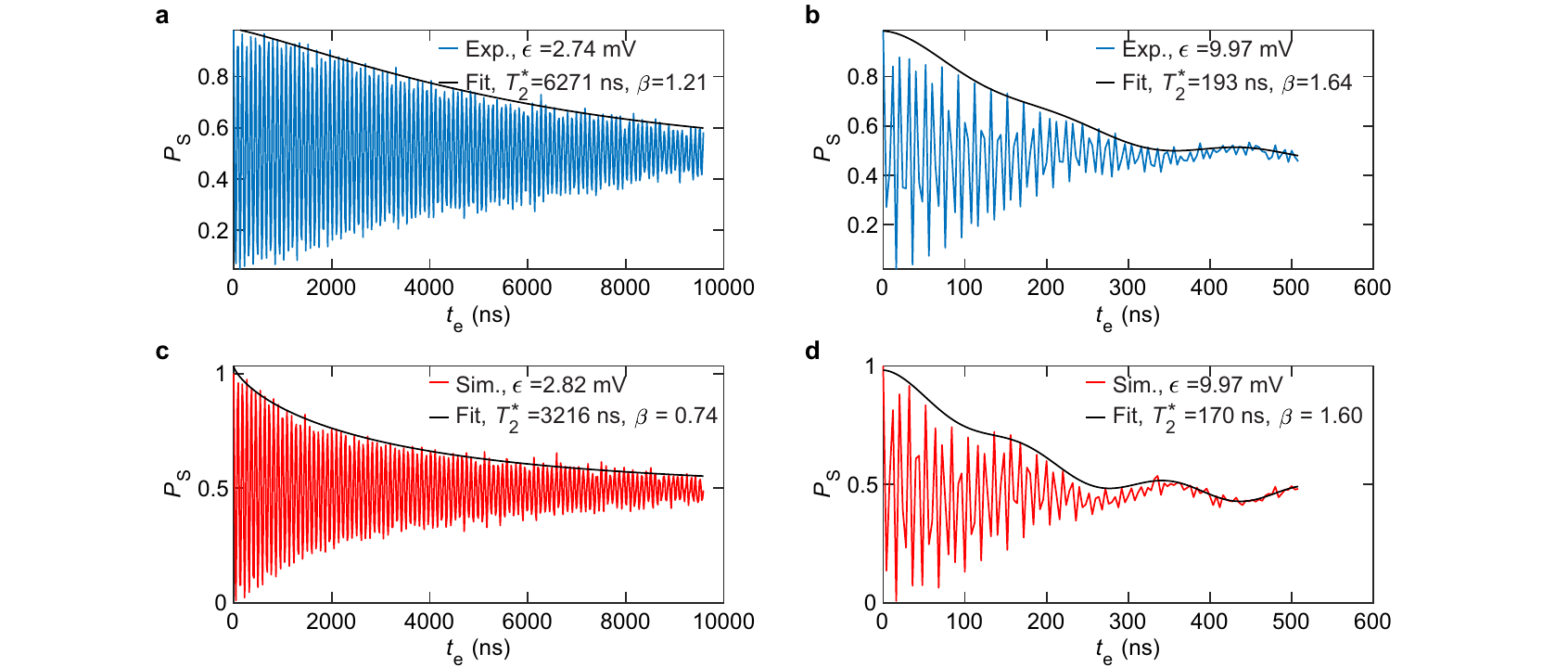}
	\caption{Measured and simulated oscillations near the third spin funnel, where the two $S^{\plus\minus}$-$T_-^{\plus\plus}$ avoided crossings have merged into one.
	\textbf{a} Measured Rabi oscillations at the avoided crossing and envelope of the corresponding fit. 
	\textbf{b} Measured Ramsey oscillations near $\eps=10$ mV and envelope of the corresponding fit.
	\textbf{c} Simulation and fit corresponding to panel \textbf{a}.
	\textbf{d} Simulation and fit corresponding to panel \textbf{b}.
	}\label{fig:sup_sim3}
\end{figure}

The simulations assume perfect state preparation and measurement. For the simulations of Figure 4 in the main text, the initial state was $\ket{S^{\plus\minus}}$, and the final state after evolution was projected via $\ket{S^{\plus\minus}}\bra{S^{\plus\minus}}$. The simulations were averaged over 100 realizations of the magnetic and electrical noise values. For the simulations of Fig.~6 in the main text and Figs.~\ref{fig:sup_sim2}-\ref{fig:sup_sim3}, the initial state was $\ket{\psi}=\frac{1}{\sqrt2}(\ket{S^{\plus\minus}}+i\ket{\tm^{\minus\minus}})$ ($\ket{\psi}=\frac{1}{\sqrt2}(\ket{S^{\plus\minus}}-i\ket{\tm^{\plus\plus}})$) for the second (third) spin funnel, and the final state after evolution was projected with $\ket{\psi}\bra{\psi}$.  The simulations were averaged over 400 realizations of the magnetic and electrical noise values. For the measurements in Fig.~6 and Figs.~\ref{fig:sup_sim2}-\ref{fig:sup_sim3}, we configured our experiments such that the total averaging time for all panels was roughly 30 minutes, and we find that the noise fluctuations described above serve to explain the coherence properties we observe, as shown in these figures. 

We hypothesize that the slight differences in $\eps$ values between the data and the corresponding simulations in Fig. 6 and Figs.~\ref{fig:sup_sim2}-\ref{fig:sup_sim3} result from small unintentional changes to the device tuning occurring in the time between calibrating the Hamiltonian parameters used in the simulations and acquiring these data, which extends over a timescale of days.
This small difference in tuning could also explain the different \stz{} oscillation frequency between Figs.~6c and 6f, and between Figs.~\ref{fig:sup_sim3}b and \ref{fig:sup_sim3}d.

Empirically, we find that the superconducting magnet in our dilution refrigerator experiences hysteresis. We notice, for example, that the spin funnels do not appear at exactly the same magnetic fields after repeated sweeps. Thus, we need to use magnetic fields in our simulations that differ by up to 20 Gauss from the setpoint values of our magnet in order to achieve good agreement with our simulations. Specifically, we use $B^z=372.06$~mT for the simulations in Fig.~4c and Fig.~6, $B^z=371.16$~mT for the simulations in Fig.~4f and Fig.~\ref{fig:sup_sim3}, and $B^z=295.4$~mT for the simulations in Fig.~4i and Fig.~\ref{fig:sup_sim2}.

\bibliography{STm_bib.bib}